\documentclass[usenatbib,lscape]{emulateapj}
\usepackage{graphicx,ifthen,url,float,color,ulem}
\newcommand {\kms}{km s$^{-1}$}

\newcommand{\scubaii}{\mbox{\sc Scuba-2}}

\newcommand {\lya}{Ly$\alpha$}
\def\ltsima{$\; \buildrel < \over \sim \;$}
\def\simlt{\lower.5ex\hbox{\ltsima}}
\def\gtsima{$\; \buildrel > \over \sim \;$}
\def\simgt{\lower.5ex\hbox{\gtsima}}

\newcommand {\um}{$\mu$m}

\def\um     {$\mu$m}
\def\ts     {\thinspace}
\def\kms    {\ifmmode{{\rm \ts km\ts s}^{-1}}\else{\ts km\ts s$^{-1}$}\fi}
\def\msol   {\ifmmode{{\rm M}_{\odot}}\else{M$_{\odot}$}\fi}
\def\lsol   {\ifmmode{{\rm L}_{\odot}}\else{L$_{\odot}$}\fi}
\def\zsol   {\ifmmode{{\rm Z}_{\odot}}\else{Z$_{\odot}$}\fi}
\def\etal   {{\rm et\ts al.}}
\def\ci     {\ifmmode{{\rm C}{\rm \small I}}\else{C\ts {\scriptsize I}}\fi}
\def\hi     {\ifmmode{{\rm H}{\rm \small I}}\else{H\ts {\scriptsize I}}\fi}
\def\hh     {\ifmmode{{\rm H}_2}\else{H$_2$}\fi}
\def\cone {\ifmmode{{\rm C}{\rm \small I}(^3\!P_1\!\to^3\!P_0)}
     \else{C\ts {\scriptsize I}{\small$(^3\!P_1\!\to\,^3\!P_0)$}}\fi}
\def\ctwo {\ifmmode{{\rm C}{\rm \small I}(^3\!P_2\!\to\,^3\!P_1)}
     \else{C\ts {\scriptsize I}{\small$(^3\!P_2\!\to\,^3\!P_1)$}}\fi}
\def\cij {\ifmmode{{\rm C}{\rm \small I}\,(^3P_i\to^3P_j)}\else{C\ts {\scriptsize I}\,{\small$(^3P_i\to^3P_j)$}}\fi}
\def\cii    {\ifmmode{{\rm C}{\rm \small II}}\else{C\ts {\scriptsize II}}\fi}
\def\tex {\ifmmode{{T}_{\rm ex}}\else{$T_{\rm ex}$}\fi}
\def\tmb {\ifmmode{{T}_{\rm mb}}\else{$T_{\rm mb}$}\fi}
\def\tkin {\ifmmode{{T}_{\rm kin}}\else{$T_{\rm kin}$}\fi}
\def\microns {\ifmmode{\mu{\rm m}}\else{$\mu$m}\fi}
\def\nhh   {\ifmmode{n({\rm H}_2)}\else{$n$(H$_2$)}\fi}

\newcommand{\msun}{{\rm\,M$_\odot$}}
\newcommand{\sfr}{{\rm\,M$_\odot$\,yr$^{-1}$}}
\newcommand{\lsun}{{\rm\,L$_\odot$}}

\newcommand{\ha}{{\rm\,H$\alpha$}}
\newcommand{\hb}{{\rm\,H$\beta$}}

\newcommand{\hd}{{\rm\,H$\delta$}}

\newcommand{\nii}{{\rm\,[N{\sc II}]}}
\newcommand{\sii}{{\rm\,[S{\sc II}]}}

\newcommand{\oii}{{\rm\,[O{\sc II}]}}

\newcommand{\oiii}{{\rm\,[O{\sc III}]}}

\shorttitle{Rest-Optical Spectra for COSMOS DSFGs}
\shortauthors{C.~M. Casey et al.}
\begin{document}

\title{Near-Infrared MOSFIRE Spectra of Dusty Star-Forming Galaxies at $0.2<z<4$}
\author{Caitlin~M.~Casey\altaffilmark{1}, Asantha~Cooray\altaffilmark{2}, Meghana
  Killi\altaffilmark{1}, Peter~Capak\altaffilmark{3}, Chian-Chou
  Chen\altaffilmark{4}, Chao-Ling Hung\altaffilmark{1}, Jeyhan Kartaltepe\altaffilmark{5}
  D.~B. Sanders\altaffilmark{5}, N.~Z. Scoville\altaffilmark{6}}
\altaffiltext{1}{Department of Astronomy, The University of Texas at Austin, 2515 Speedway Blvd Stop C1400, Austin, TX 78712}
\altaffiltext{2}{Department of Physics and Astronomy, University of California, Irvine, Irvine, CA 92697}
\altaffiltext{3}{Infrared Processing and Analysis Center (IPAC), 1200 E. California Blvd., Pasadena, CA 91125}
\altaffiltext{4}{European Southern Observatory, Karl-Schwarzschild-Strasse 2, 85748 Garching bei M\"{u}nchen, Germany}
\altaffiltext{5}{School of Physics and Astronomy, Rochester Institute of Technology, 54 Lomb Memorial Drive, Rochester, NY 14623}
\altaffiltext{6}{Institute for Astronomy, University of Hawai'i at Manoa, 2680 Woodlawn Dr, Honolulu, HI 96822}
\altaffiltext{7}{California Institute of Technology, 1216 East California Boulevard, Pasadena, CA 91125}
\label{firstpage}

\begin{abstract}
We present near-infrared and optical spectroscopic observations of a
sample of 450\um\ and 850\um-selected dusty star-forming galaxies
(DSFGs) identified in a 400\,arcmin$^2$ area in the COSMOS field.
Thirty-one sources of the 114 targets were
spectroscopically confirmed at $0.2<z<4$, identified primarily in the
near-infrared with Keck MOSFIRE and some in the optical with Keck LRIS
and DEIMOS.  The low rate of confirmation is attributable both to high
rest-frame optical obscuration in our targets and limited sensitivity
to certain redshift ranges.  The median spectroscopic
  redshift is $\langle z_{\rm spec}\rangle=1.55\pm0.14$, comparable to
  $\langle z_{\rm phot}\rangle=1.50\pm0.09$ for the larger parent DSFG
  sample; the median stellar mass is
  $(4.9^{+2.1}_{-1.4})\times10^{10}\,$\msun, star-formation rate is
  160$\pm$50\,\sfr, and attenuation is $A_{\rm V}=5.0\pm0.4$.  The
high-quality photometric redshifts available in the COSMOS field allow
us to test the robustness of photometric redshifts for DSFGs.  We find
a subset (11/31$\approx35$\%) of DSFGs with inaccurate ($\Delta
z/(1+z)>0.2$) or non-existent photometric redshifts; these have very
distinct spectral energy distributions from the remaining DSFGs,
suggesting a decoupling of highly obscured and unobscured components.
We present a composite rest-frame 4300--7300\AA\ spectrum for DSFGs,
and find evidence of 200$\pm$30\,\kms\ gas outflows.  Nebular line
emission for a sub-sample of our detections indicate that hard
ionizing radiation fields are ubiquitous in high-$z$ DSFGs, even more
so than typical mass or UV-selected high-$z$ galaxies. We also confirm
the extreme level of dust obscuration in DSFGs, measuring very high
Balmer decrements, and very high ratios of IR to UV and IR to
\ha\ luminosities.  This work demonstrates the need to broaden the use
of wide bandwidth technology in the millimeter to 
  spectroscopically confirm larger samples of high-$z$ DSFGs, as the
difficulty in confirming such sources at optical/near-infrared
wavelengths is exceedingly challenging given their obscuration.
\end{abstract}


\section{Introduction}

The most challenging observational hurdle in the study of extremely
obscured galaxies has been obtaining accurate spectroscopic redshifts
\citep{chapman03a,swinbank04a,chapman05a}.  These redshifts present a
major bottleneck in understanding the physics of these extreme
star-formers, and limit our ability to test evolutionary models
derived from simulations \citep[e.g.][]{lacey15a,narayanan15a}.
Gaining insight into the physics of Dusty Star-Forming Galaxies
(DSFGs) is of crucial importance in constraining massive galaxy
formation and the buildup of stellar mass at early times, when these
systems dominate cosmic star-formation \citep*[see reviews of][]{casey14a,blain02a}.

Since the first identification of submillimeter-luminous galaxies with
the Submillimeter Common User Bolometric Array (SCUBA) instrument
\citep{smail97a,hughes98a,barger98a}, the characterization of DSFGs
has been a top priority of extragalactic astrophysics.  The large
beamsize of SCUBA at the James Clerk Maxwell Telescope (JCMT) made the
immediate identification of DSFGs' multiwavelength counterparts
difficult.  However, the correlation between radio and FIR luminosity
in starburst galaxies \citep{helou85a,condon92a} provided a means of
using radio interferometric positions to identify the likely source of
FIR/submillimeter emission \citep{yun01a,chapman04a,chapman05a}.  Yet,
redshift identification was still a challenge.  Optical/near-infrared
photometric redshifts are often out of reach for these highly obscured
galaxies, and obtaining spectroscopic redshifts required the most
sensitive optical and near-IR spectrographs on the largest telescopes,
such as the Low Resolution Imaging
Spectrometer (LRIS) on Keck \citep{oke95a}.  With star-formation rates
exceeding 100\sfr, only $\simlt$0.5\%\ of 
  rest-frame UV/optical starlight, and optical/ultraviolet emission
features, are unobscured by dust \citep{howell10a,casey14b}.

The most comprehensive survey of spectroscopic redshifts for DSFGs was
presented in \citet{chapman05a}.  They present 75 redshifts for
submillimeter galaxies (SMGs\footnote{In this paper and in recent
  years, we have adopted the term DSFG as a more general name for
  SMGs.  SMGs have been formally defined in the past as having an
  850\um\ flux density in excess of $\sim$2\,mJy, while DSFG
  represents any galaxy directly detected at wavelengths
  $\sim$70\um--2\,mm with current and past facilities, with the
  exclusion of ALMA which is much more sensitive.}) peaking at
$z\sim2.5$.  This sample was the primary sample pursued for follow-up,
from CO molecular gas studies
\citep{greve05a,tacconi06a,tacconi08a,engel10a,bothwell10a,bothwell13a,casey11a},
to rest-frame optical follow-up
\citep{swinbank04a,menendez-delmestre13a,alaghband-zadeh12a}, to
mid-infrared follow-up from {\it Spitzer} IRS
\citep{pope08a,menendez-delmestre09a,coppin10a}, and high-resolution
radio continuum mapping \citep{biggs08a,casey09b}.
These studies pointed out that the 850\um-selected DSFG population is
primarily made up of major galaxy mergers at the tip of the luminosity
function and could be more heterogeneous in triggering mechanism at
intermediate luminosities.  Nevertheless, they exhibit exceptional
star-formation rates which are predominantly short-lived
\citep{bothwell13a,swinbank14a} and result in the formation of the
most massive galaxies in the Universe.

\begin{table*}
\caption{Details of MOSFIRE Observations}
\centering
\begin{tabular}{cccccccccc}
\hline\hline
{\sc Slitmask} & {\sc Obs.} & {\sc Mask Position} & {\sc Exp. Time} & {\sc Exp. Time} & N$_{\rm S2}$ & N$_{\rm all}$ & {\sc N($z$}) & {\sc \%-DSFGs conf.} \\
   {\sc Name}  &  {\sc Date}    &           & {\sc $K$-band} [s] & {\sc $H$-band} [s] & & & {\sc DSFGs} & [\%] \\
\hline
cosm1 & 21-Dec-12 & 10:00:10.25$+$02:19:44.71 & 2880 & 1920 & 12 & 25 & 3 & 25\% \\ 
cosm2 & 21-Dec-12 & 10:00:18.94$+$02:24:28.70 & 2520 & 1200 & 10 & 24 & 4 & 40\% \\ 
cosm3 & 21-Dec-12 & 10:00:23.39$+$02:30:37.29 & 2520 & 720  & 12 & 24 & 1 & 8\% \\ 
cosm4 & 31-Dec-13 & 10:00:31.91$+$02:20:13.00 & 3600 & 2880 & 16 & 21 & 9 & 56\% \\ 
cosm5 & 19-Jan-14 & 10:00:51.13$+$02:20:57.92 & 3240 & 1440 & 10 & 25 & 4 & 40\% \\ 
cosm6 & 31-Dec-13 & 10:00:01.15$+$02:24:35.91 & 3600 & 1920 & 13 & 23 & 3 & 23\% \\ 
cosm7 & 31-Dec-13 & 09:59:51.54$+$02:21:07.69 & 2880 & 1320 & 14 & 25 & 6 & 43\% \\ 
cosm8 & 19-Jan-14 & 10:00:15.16$+$02:16:38.68 & 3600 & 2400 & 15 & 26 & 4 & 27\% \\ 
cosm9 & 19-Jan-14 & 10:00:00.67$+$02:28:15.20 & 3600 & 2880 & 16 & 24 & 5 & 31\% \\ 
cosm10 & 19-Jan-14 & 10:01:00.05$+$02:24:59.20 & 2880 & $-$ & 10 & 24 & 0 & 0\% \\ %
{\sc TOTAL} &      &                          &      &      & 102 &   & 31 & 29.8\% \\
\hline\hline
\end{tabular}
\label{tab:obs}

{\small {\bf Table Description.}
Our MOSFIRE observations consisted of 10 slitmask configurations, nine
of which were observed in both $H$- and $K$-band.  The exposure time and
number of targets per mask varied as a function of real-time observing
decisions based on accessibility of the field, intermittent cloud
cover, and higher-priority vs. lower-priority targets.  The number of
{\sc Scuba-2} sources that were targeted per slitmask is given in $N_{\rm
  S2}$ while the total number of targets on that slitmask is given in
$N_{\rm all}$.  The number of DSFGs that were spectroscopically
confirmed on the slitmask, a subset of $N_{\rm S2}$, is given in the
N($z$) DSFGs column.  The percentage of DSFGs with spectroscopic
confirmations on the given slitmask is given in the last column,
ranging from 0--56\%, and averaging 29.8\%.
}
\end{table*}

\begin{figure}
\includegraphics[width=0.49\columnwidth]{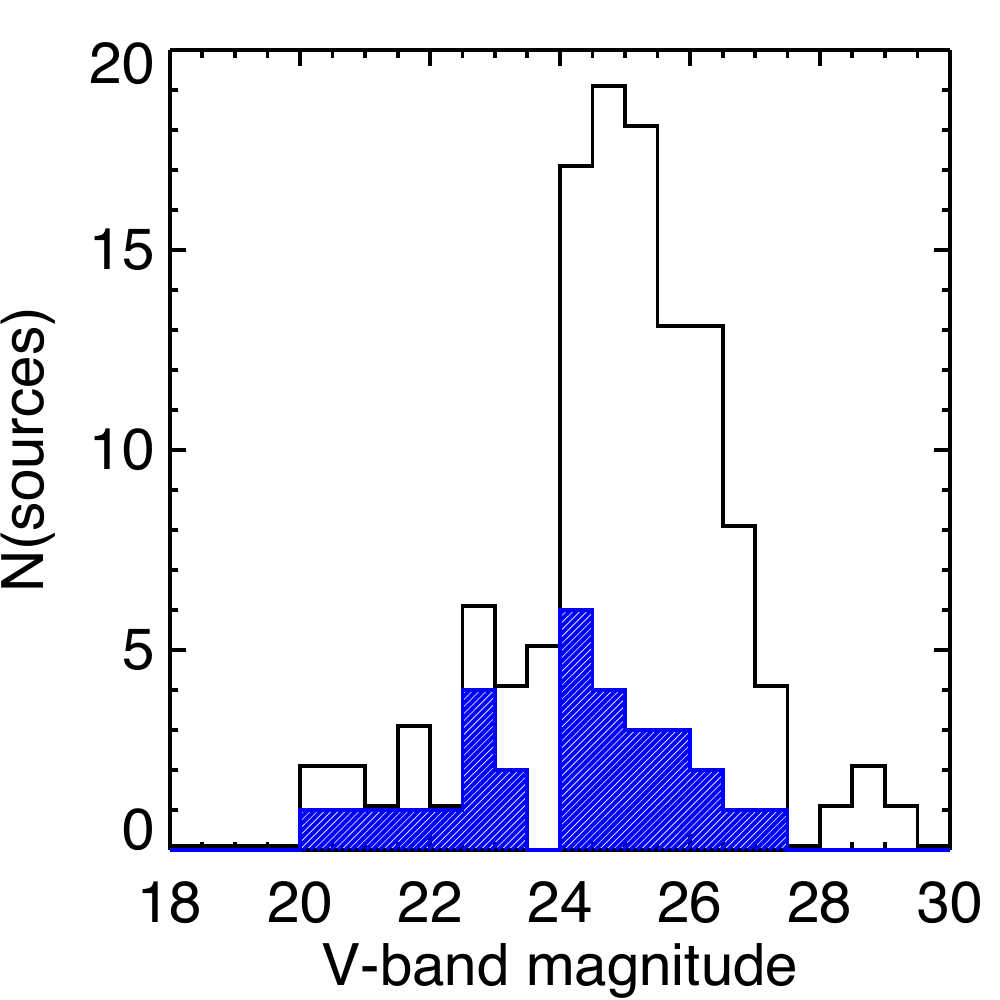}
\includegraphics[width=0.49\columnwidth]{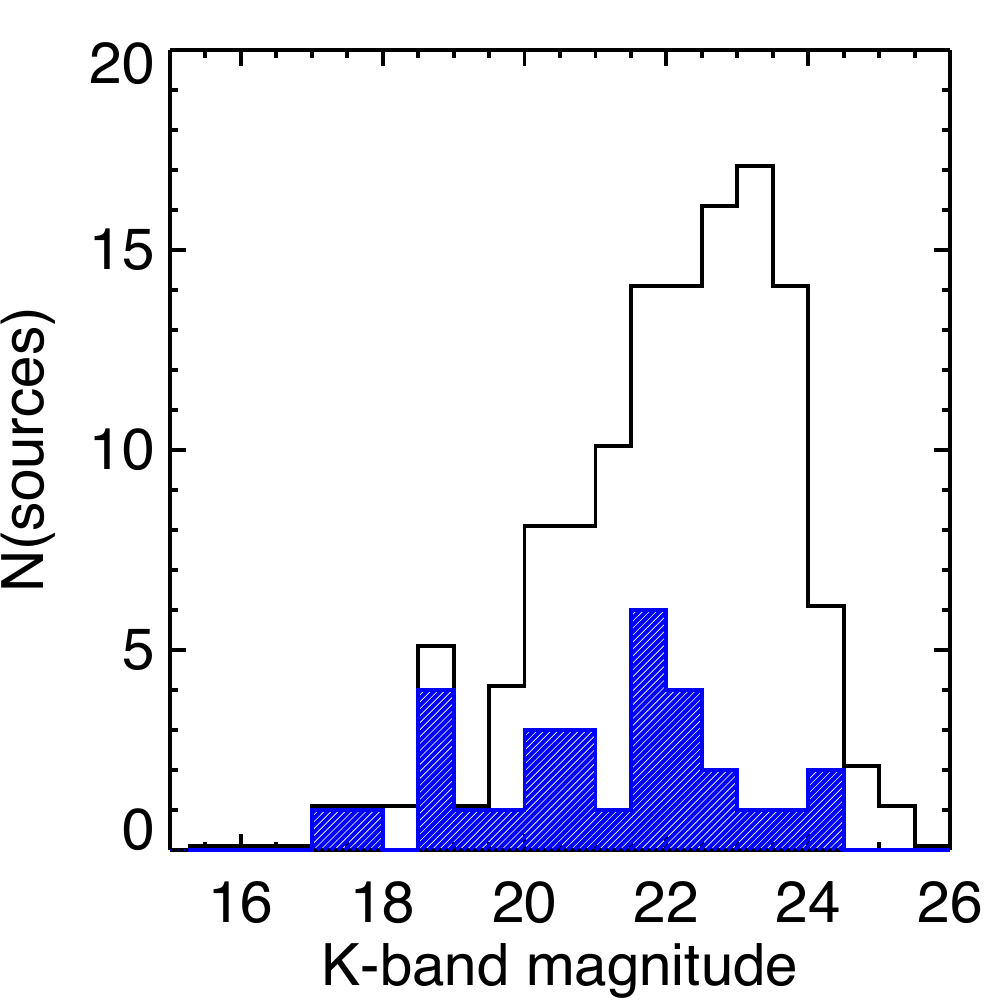}
\includegraphics[width=0.49\columnwidth]{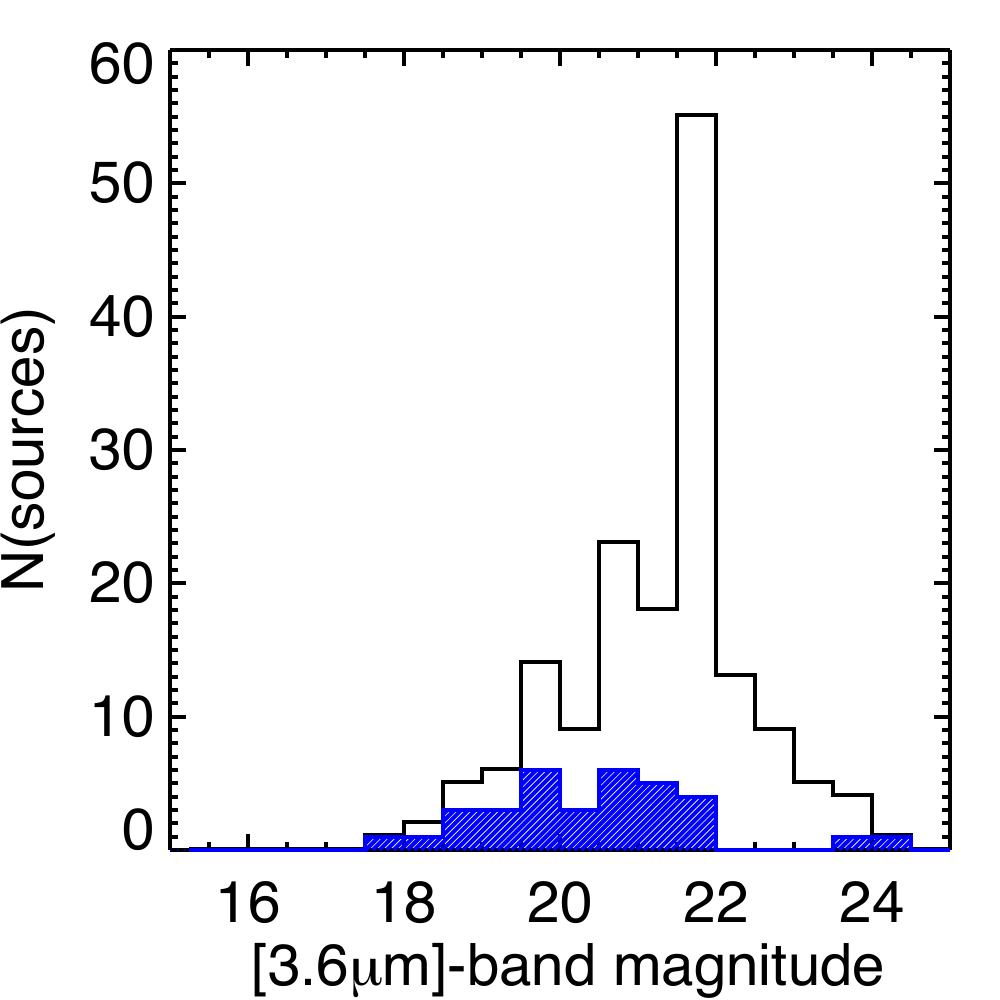}
\includegraphics[width=0.49\columnwidth]{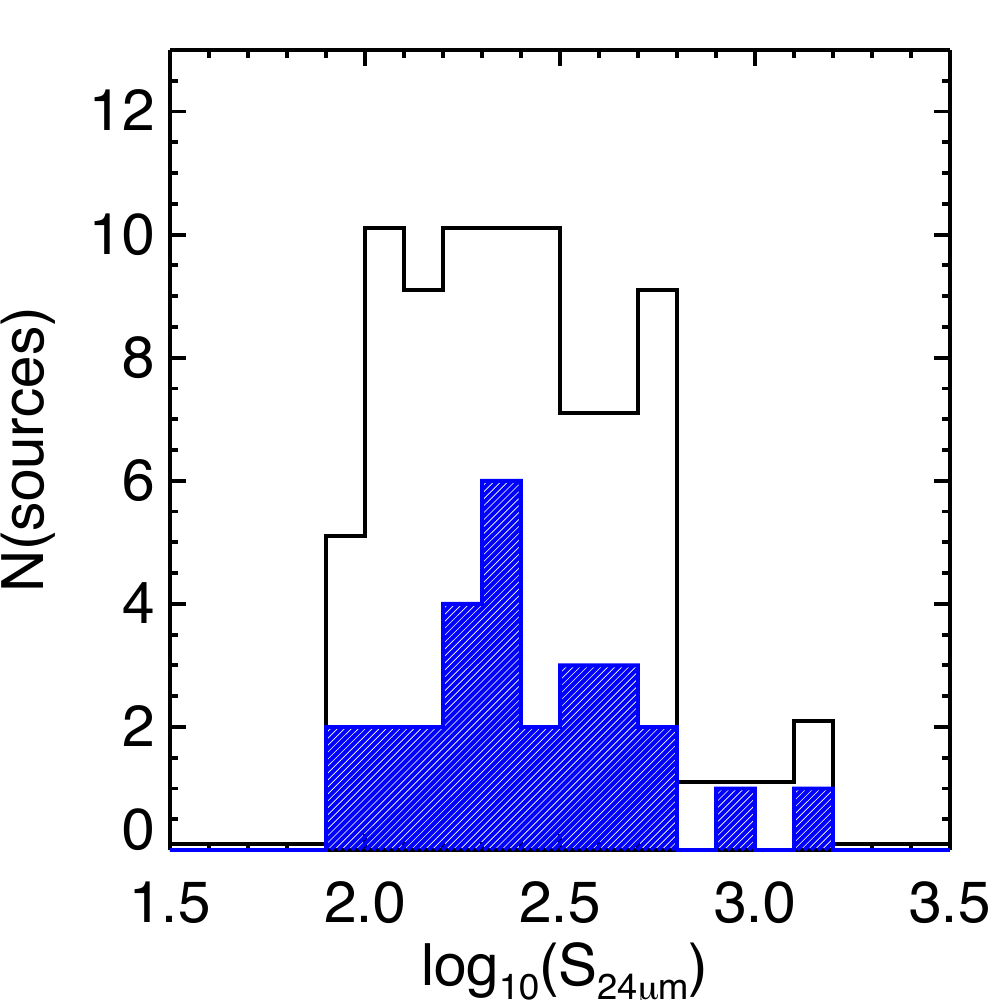}
\caption{Comparison of sources' magnitudes in V-, K- and
    3.6\um\ bands and flux densities at 24\um\ between the parent
    sample of 165 \scubaii\ sources and those with spectroscopic
    redshifts described in this paper.}
\label{fig:parent}
\end{figure}

Though those with spectroscopic identification have proved very
valuable to DSFG research, there are still few constraints on the
$\sim$50\%\ of the population of DSFGs which have not been detected
via Lyman-$\alpha$ emission or other rest-frame ultraviolet tracers.
Do they lie at higher redshifts?  Are they significantly more obscured
by dust?  Are their physical triggering mechanisms different from
those with detectable UV/optical emission features?  To address the
unknown selection bias the UV/optical (and radio detected) sub-sample
may place on our interpretation of DSFGs, the ALESS Survey has set out
to meticulously characterize an unbiased population of 870\um-selected
DSFGs with rest-frame UV/optical spectroscopy and submillimeter
interferometry from the Atacama Large Millimeter Array
\citep[ALMA][]{weis09a,wardlow11a,hodge13a}.  Spectroscopic success
rates for DSFGs with known interferometric positions is $<$50\%\ using
optical/near-infrared \citep{danielson16a}.  While ALMA has ushered in
a new era of spectroscopically confirming DSFGs via direct emission of
CO and [C{\sc II}] at long wavelengths, this unfortunately is still
prohibitively expensive for large samples of unlensed galaxies
\citep[c.f.][]{vieira13a}.
\begin{figure}
\includegraphics[width=0.99\columnwidth]{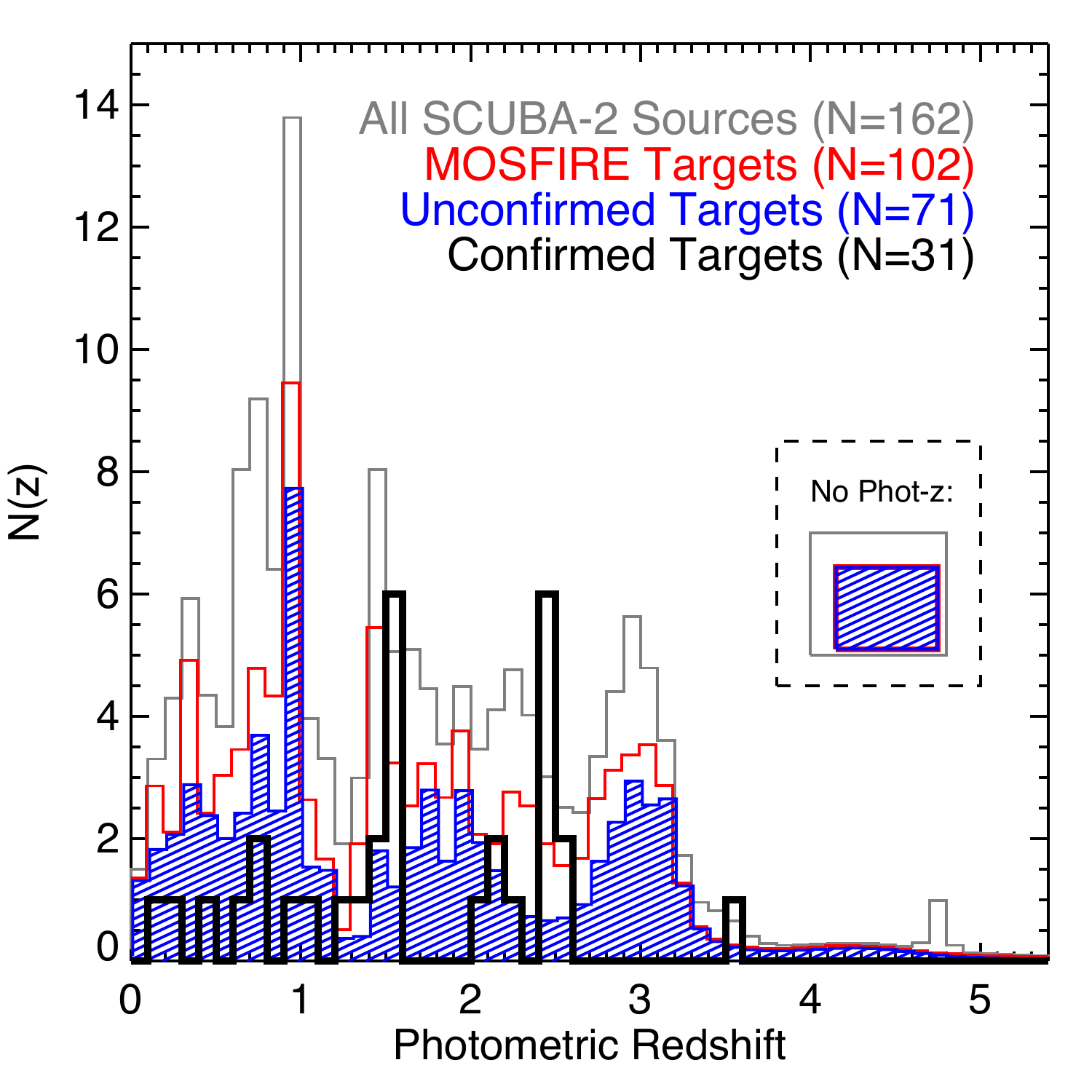}
\caption{The distribution in photometric redshifts for the parent
  sample of SCUBA-2 sources in COSMOS \citep[selected from][;
    gray]{casey13a}, the sample targeted by MOSFIRE (red, primarily
  determined by spatial configuration on the sky with respect to
  slitmasks), the sample which has been targeted but is {\it not}
  spectroscopically confirmed (blue hashed region), and those that
  have been confirmed (black distribution).  In this depiction, each
  galaxy's photometric redshift is represented by an asymmetric
  Gaussian with area under the curve equal to one.  This more
  accurately portrays the photometric redshift constraints than a
  simple histogram.  The total number of sources depicted here is 162
  (total sample), 102 (targets selected for MOSFIRE follow-up), 71
  (unconfirmed sources), and 31 (confirmed sources).  We include
  targets without photometric redshift fits (inside dashed box) for
  comparison.}
\label{fig:zphotdist}
\end{figure}

We have embarked on an independent spectroscopic follow-up campaign of
DSFGs selected at 450\um\ and 850\um\ using the SCUBA-2 instrument in
the COSMOS field \citep{casey13a}.  Due to the extreme obscuration
present in DSFGs at rest-frame UV wavelengths, our spectroscopic
campaign has primarily focused on rest-frame optical features
detectable with the superbly sensitive MOSFIRE instrument on Keck
\citep{mclean12a}. Though only a subset of DSFGs in this sample have
interferometric measurements from ALMA and the Plateau de Bure
Interferometer, 62\%\ of the DSFGs in the sample are detected at
450\um, which has a significantly smaller beamsize (7\arcsec) than
850\um\ (15\arcsec) or 870\um\ on LABOCA (19\arcsec).

Here we present the results of this spectroscopic survey. In
\S~\ref{sec:obs} we describe the spectroscopic data.  In
\S~\ref{sec:photz} we present an analysis of the sample's redshifts,
contrasting photometric and spectroscopic identifications, and we
discuss individual source characteristics in \S~\ref{sec:individual}.
In \S~\ref{sec:stack} we present a composite rest-frame optical
spectrum comprised of 20 $z=1.26-2.55$ DSFGs, and we compare to
previous composites presented for DSFG samples.  In \S~\ref{sec:agn}
we present an analysis of the sample's Active Galactic Nuclei (AGN)
using X-ray coverage in the field.  In \S~\ref{sec:bpt} we present
nebular line diagnostics for the handful of galaxies and use them to
analyze possible physical drivers of line emission.  In
\S~\ref{sec:ha} we discuss the H$\alpha$ luminosities of the DSFGs we
sample, in relation to their SFRs measured at long wavelengths.  In
\S~\ref{sec:conclusions} we present our conclusions.  Throughout, we
assume a $\Lambda$\,{\sc CDM} cosmology with $H_{\rm
  0}$=71\,km\,s$^{-1}$\,Mpc$^{-1}$ and $\Omega_{\rm m}$=0.27
\citep{hinshaw09a}. We also assume a Chabrier initial mass function
\citep[IMF;][]{chabrier03a}.

\section{Observations \&\ Data}\label{sec:obs}

The parent sample for spectroscopic follow-up we use in
  this analysis is the full set of 165 \scubaii-identified DSFGs in
  COSMOS, as summarized in \citet{casey13a}.  The 1-$\sigma$
  sensitivity of the \scubaii\ maps used were 4.13\,mJy and 0.80\,mJy
  at 450\um\ and 850\um\ respectively.  Of the 165 independent sources
  identified, 78 were detected above 3.6$\sigma$ at 450\um\ and 99
  above the same threshold at 850\um.  And additional eight sources
  were identified at marginal $3<\sigma<3.6$ significance at both
  450\um\ and 850\um.  About 2/3 of the sample are detected in both
  450\um\ and 850\um\ bands.  With positional uncertainties of
  1--2.5\,\arcsec, optical/near-infrared counterparts were identified
  directly from submillimeter positions weighted by a $p$-value.
  Sixty-one percent of the entire sample has `high confidence' OIR
  counterparts identified using this method, with
  $<$5\%\ contamination; the remaining 39\%\ of the parent sample is
  less well matched.  The targets which were observed
  spectroscopically were not biased towards either secure or low
  confidence populations. We discuss the relationship of the
  spectroscopic sample to the parent sample more in \S~\ref{sec:photz}
  and refer the reader to \citet{casey13a} for more details.

Our near-infrared spectroscopic follow-up campaign was carried out on the
Keck 1 MOSFIRE instrument on 2012 December 21, 2013 December 31 and
2014 January 19.  Observing conditions for these nights were all
favorable, with clear skies and 0\arcsec.5--0\arcsec.7 seeing.  We
observed ten MOSFIRE masks, all designed in the MAGMA mask design
software package for MOSFIRE in $K$-band, and nine of the same masks in
$H$-band all using an ABBA 1\arcsec.5 nod pattern.  Individual exposures
in $H$-band were 120\,s while individual exposures in $K$-band were
180\,s.  Observation details are given in Table~\ref{tab:obs}.  Total
on-source integration times varied from 720--2880\,s in the $H$-band and
2520--3600\,s in the $K$-band.  No $Y$-band or $J$-band observations were
taken.  To maximize the number of high priority targets per mask, we
made some minor sacrifices in spectral coverage; on average each
primary target had complete wavelength coverage from 1.45--1.75\um\ in
$H$-band and from 1.92--2.40\um\ in $K$-band.  We used the MOSPY Data
Reduction Pipeline (DRP) package for spectral reduction.  In total,
102 DSFG targets were observed with MOSFIRE.

We also present observations obtained with the DEIMOS instrument on
Keck II, following-up the same SCUBA-2 selected sample at optical
wavelengths.  DEIMOS observations were carried out on 10-11 December
2012, 07 February 2013, and 28 October 2014.  Conditions were
suboptimal on 10-11 December 2012 with intermittent cloud cover and
1.1-1.3\arcsec\ seeing.  Conditions on 07 February 2013 were poor and
completely weathered out during the LST when the COSMOS field was
accessible.  We used the 600 lines\,mm$^{-1}$ grating with a
7200\,\AA\ blaze angle (resulting in dispersion of 0.65\,\AA) and the
GG455 filter to block out higher-order light.  Wavelength coverage
varied with sources' positions on the slit-mask, averaging
4800--9500\,\AA.  Twelve sources were observed with DEIMOS that were
not observed with MOSFIRE, but the majority of DEIMOS observations
overlapped with MOSFIRE targets.

\begin{figure}
\vspace{-1.5cm}
\includegraphics[width=0.99\columnwidth]{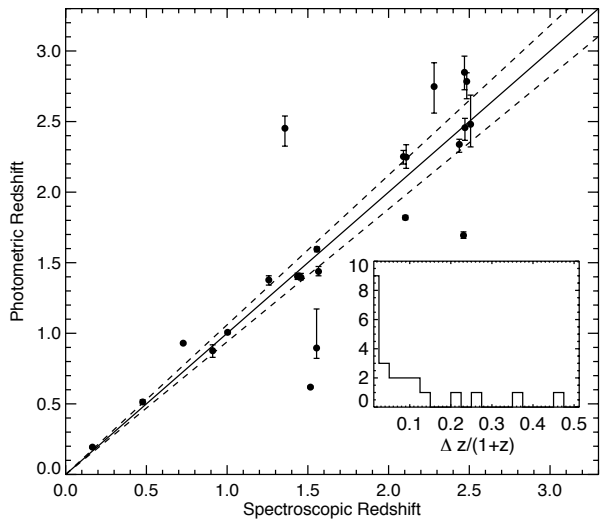}
\vspace{-1.5cm}
\caption{Comparison of spectroscopic and photometric redshifts for
  SCUBA-2 galaxies in COSMOS.  Dashed lines show the $\Delta z$/(1+z)
  = 0.05 level of accuracy which is the average of this dataset.  Note
  this comparison excludes five sources which lack photometric redshift
  estimates.  On occasion there was a slight positional offset between
  photometric and spectroscopic sources, yet no larger than
  0.\arcsec4. The inset plot shows a histogram of $\Delta z$/(1+z).
  As discussed in the text \S~\ref{sec:photz}, the outliers have
  significantly higher IR luminosities than their optical/near-infrared
  emission would imply (also see Figure~\ref{fig:magphys}).}
\label{fig:zspeczphot}
\end{figure}
All spectra, from MOSFIRE and DEIMOS, were extracted using the {\sc
  iraf} routine {\sc apall}.  Apertures were set interactively,
depending on the seeing and whether or not the galaxy is spatially
resolved on the slit, and background is subtracted using
simultaneously-fit sky apertures.  Sources are corrected for a trace
only when continuum emission is of significant signal-to-noise across
the entire spectral region, otherwise a fixed aperture is
assumed. Optimal extraction is set with variance weighting.

 Figure~\ref{fig:parent} compares the characteristics of
  the parent sample of SCUBA-2 galaxies with the sources which have
  been spectroscopically confirmed.  While one might expect the
  strongest correlation with $K$-band magnitude (as this is the
  wavelength regime of the MOSFIRE spectroscopy), we see a relatively
  uniform distribution in magnitude for confirmed sources.  Though
  this does suggest a lower fraction of confirmations at fainter
  magnitudes, the distributions are not statistically inconsistent
  with the parent sample.  The distribution in 24\um\ flux densities
  is similarly uncorrelated with spectroscopic confirmation.  The
  statistics for 1.4\,GHz radio emission is similar, yet has fewer
  sources than our 24\um\ distribution.  We attribute the lack of
  correlation to our spectroscopic follow-up strategy which was
  intended to not be biased with relative near-infrared or optical
  luminosity.  The selection of targets was focused on maximizing the
  number of SCUBA-2 targets per mask and so represents a spatial
  selection on 6\arcmin\ scales, not noticeably biased with respect to
  sample characteristics.

We also make use of the extensive multi-wavelength ancillary data
available in the COSMOS field \citep{capak07a}, including over 30
optical and near-infrared photometry and associated photometric
redshifts \citep{ilbert13a,lagache15a}, spectroscopic redshifts
\citep{lilly09a}, X-ray imaging from {\it Chandra} \citep{civano12a},
      {\it Herschel} PEP/PACS and HerMES/SPIRE 100--500\um\ catalogs
      \citep{lutz11a,oliver12a,lee13a}, and deep radio continuum
      mapping at 1.4\,GHz \citep{schinnerer07a} and 3.0\,GHz
      \citep{smolcic16a}.  It should be emphasized that the quality of
      photometry in the COSMOS field is superb and provides a unique
      opportunity to test the reliability of high quality photometric
      redshifts for highly obscured galaxies.  Seven of
        our spectroscopically-confirmed sources also have 1.1\,mm ALMA
        dust continuum maps available from the ALMA archive, programs
        \#2013.1.00118.S, 2013.1.00151.S, and 2011.1.00539.S.  Five of
        the seven sources were correctly identified using
        multiwavelength counterparts (71\%), as we discuss further in
        \S~\ref{sec:dsfgs}.  Those that were incorrectly identified
        were still detected, but submm counterparts were
        misidentified.  The ALMA 1.1\,mm maps range in RMS from
        0.08--0.15\,mJy/beam, sources are detected with a median
        S/N=10, and the median ratio between 850\um\ flux density and
        1.1\,mm flux density is 3.9$\pm$0.6.

\section{Photometric Redshift Analysis}\label{sec:photz}

Since COSMOS has some of the most precisely constrained photometric
redshifts for even the faintest sources, we are able to assess the
quality of those photometric redshifts for highly obscured sources
like DSFGs \citep[e.g.][]{wardlow11a}.  While our original sample in
\citet{casey13a} made use of the then-current COSMOS team photometric
redshift catalog \citep{ilbert13a}, in this analysis we update those
photometric redshifts to the most current catalog \citep{laigle16a}.

Figure~\ref{fig:zphotdist} shows the distribution of photometric
redshifts (phot-$z$'s) for the parent sample of 162 SCUBA-2 sources in
the COSMOS field.  This includes both 450\um\ and 850\um-selected
sources, as well as `marginal' detections identified at both
wavelengths (denoted with a preceding `m'), at a signal-to-noise of
$3<\sigma<3.6$ in both filters \citep[see][]{casey13a}.  It should be
noted that this redshift distribution is significantly different than
the distribution of either the 450\um\ or 850\um\ sources, who peak at
$\langle z_{\rm 450}\rangle=1.95\pm0.19$ and $\langle z_{\rm
  850}\rangle=2.16\pm0.11$ respectively.  The median of this parent
distribution is $\langle z\rangle=1.46\pm0.14$, similar to the results
of \citet{roseboom13a}.  We attribute this to the inclusion of the
marginally detected 450\um\ and 850\um\ sources to our spectroscopic
follow-up sample: these sources are less submillimeter luminous and so
are likely at lower redshifts \citep[see also][]{bethermin15a}.

Figure~\ref{fig:zphotdist} also shows the phot-z distribution for
sources targeted by our MOSFIRE campaign (102 sources), and those
which have not been confirmed (71 sources) and confirmed (31
sources).   
Those sources which have been spectroscopically confirmed do trend
towards redshift ranges which are well-suited to the identification of
strong emission and absorption features in the observed $H$- and
$K$-bands (i.e. $z\sim1.2--1.7$ and $z\sim2.0-2.5$). Many sources with
photometric redshifts outside of this range did show continuum
emission in our $H$- and $K$-band observations, but the lack of
discernable emission or absorption feature makes redshift
identification not possible.  sSources with photometric redshifts in
our optimum range without spectroscopic confirmations are on average 1
magnitude fainter in the near-infrared than those that have been
spectroscopically confirmed.  The lack of spectroscopic confirmation,
however, is probably attributable to a mix of 
 luminosity, geometry, \begin{figure*}
\includegraphics[width=1.99\columnwidth]{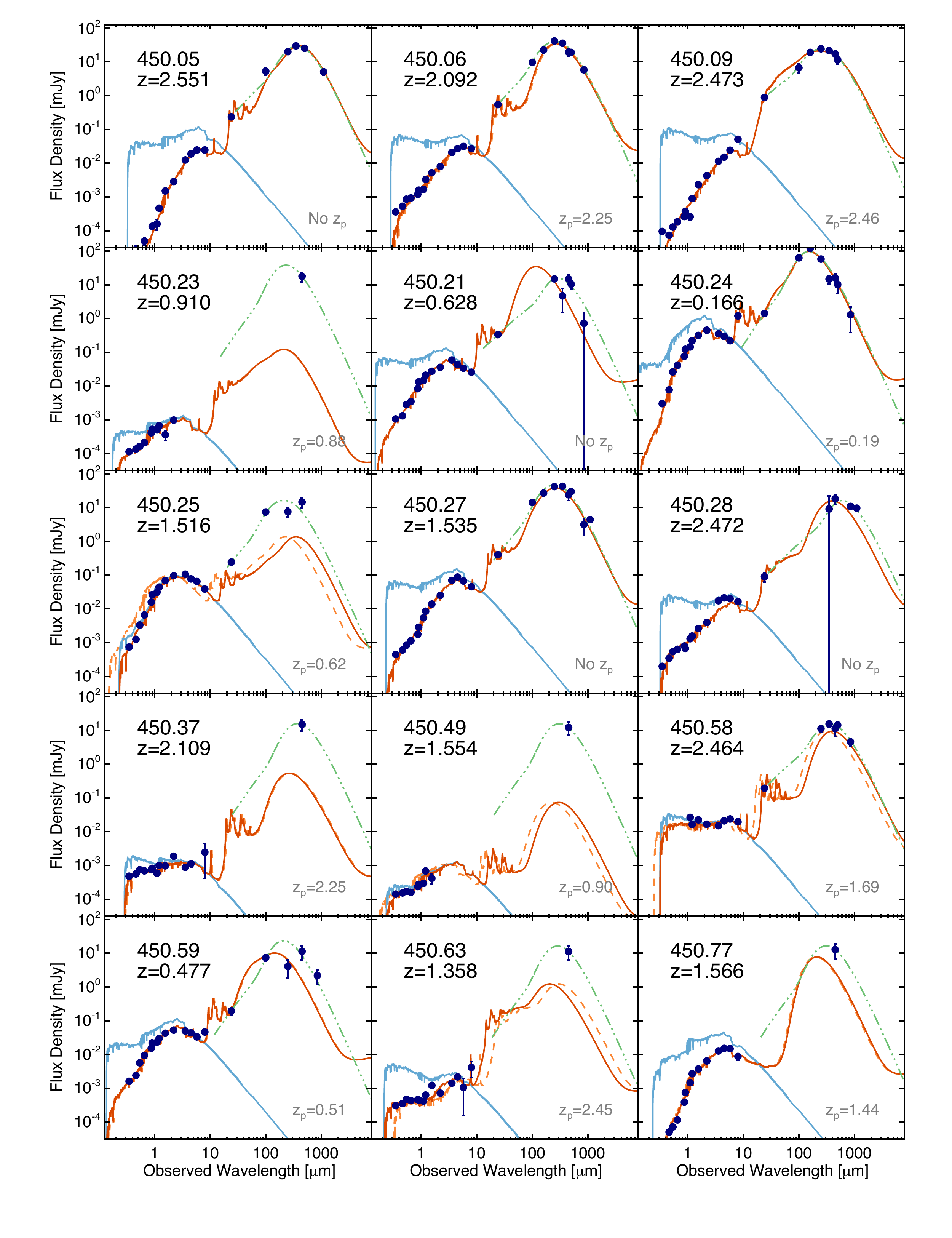}
\caption{The UV through submillimeter photometry for the DSFGs in our
  sample with confirmed spectroscopic redshifts; spectral energy
  distribution fits come from \citet{bruzual03a} models using the {\sc
    Magphys} energy balance code \citep{da-cunha08a}.  450.14 is
  excluded due to its multiple redshift solution, and thus blended
  photometry. The corresponding unobscured SED is shown in light blue.
  In addition, we fit simple modified blackbodies plus mid-infrared
  powerlaws to the photometry longward of 20\um\ (green dot-dashed
  lines) to compare to the best fit {\sc Magphys} SED output.  While
  often consistent, several galaxies show a strong disconnect between
  best-fit simple FIR SED and an SED which takes the rest-frame UV and
  optical into account, indicating a complete decoupling of
  UV-emitting and submillimeter-emitting regions.}
\label{fig:magphys}
\end{figure*}
\clearpage
\begin{center}
\includegraphics[width=1.99\columnwidth]{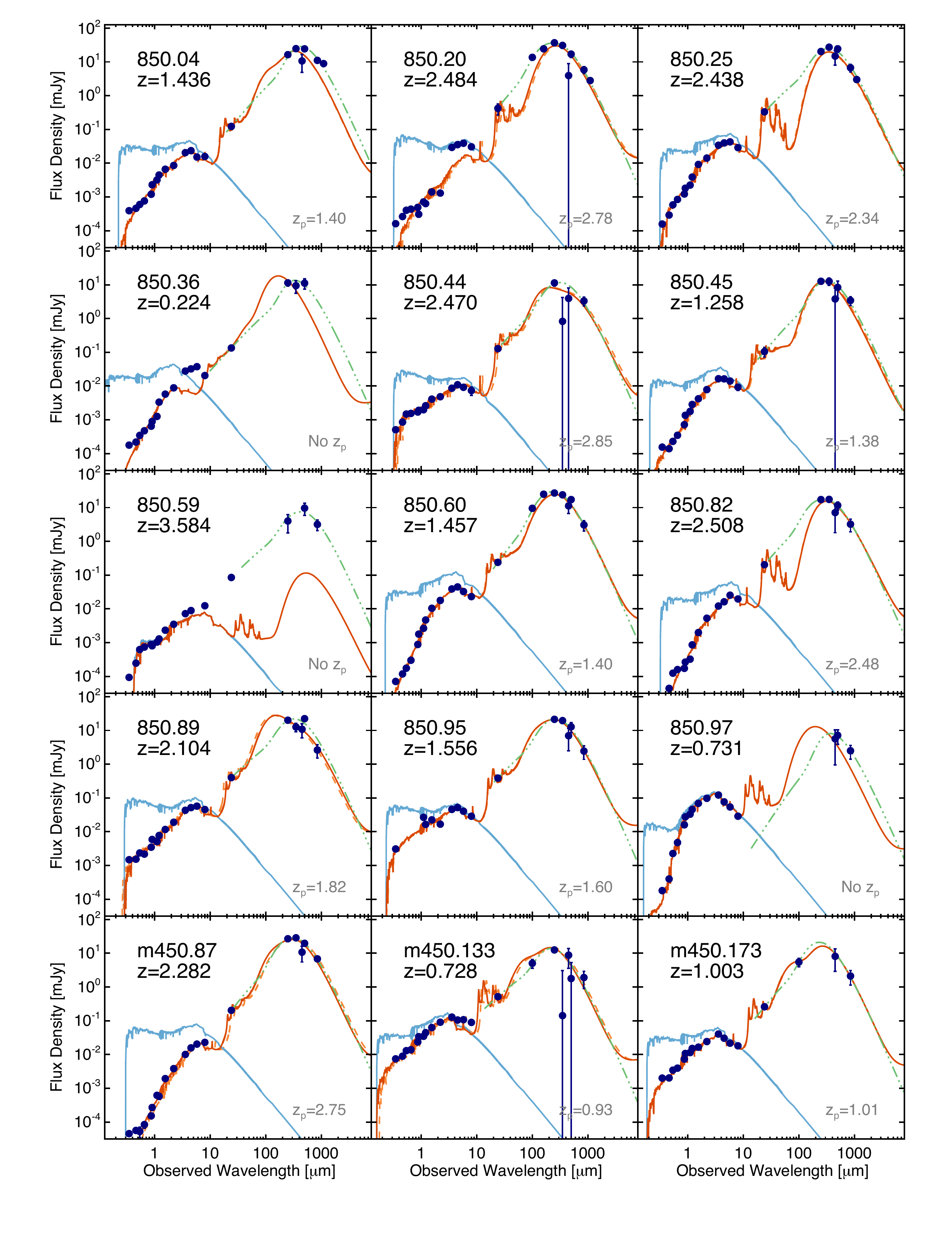}
\centerline{{\small Figure~\ref{fig:magphys} --- continued.}}
\end{center}
\clearpage
 \noindent and inaccurate photometric redshifts for
  dust-obscured sources.


Figure~\ref{fig:zspeczphot} illustrates the correlation of photometric
redshift with spectroscopic redshift.  For confirmed DSFGs with
photometric redshifts in COSMOS, these photometric redshifts prove to
be fairly reliable, with an average $\Delta z/(1+z)=$0.05, only
moderately worse than the average for `normal,' less obscured
galaxies.  This is encouraging, as it demonstrates that high-quality
photometric redshifts and a broad range of templates that include a
range of star-formation histories can accurately reflect a DSFG's
redshift. However, the distribution of $\Delta z/(1+z)$ is broadened
relative to normal galaxies with many examples of catastrophic
failures: 4 sources have $\Delta z/(1+z)>0.2$ and 7 lack any
photometric redshift solution.

What gives rise to such inaccuracies in DSFGs' photometric
redshifts in some cases but not others?  We investigate the possible
causes of such inaccuracies by fitting spectral energy distributions
(SEDs) to the UV through millimeter photometry using the MAGPHYS code
\citep{da-cunha08a}.  We use the updated version \citep{da-cunha15a},
equipped to fit SEDs of high-redshift, potentially burst-dominated
galaxies like DSFGs.  The  SED fits are shown in
Figure~\ref{fig:magphys}.

Geometric effects are the most likely culprit causing significant
photometric inconsistencies, leading to poorer photometric redshift
estimation.  For example, 
we identify five galaxies with MAGPHYS fits that are not well matched
from UV through IR in Figure~\ref{fig:magphys}: 450.23, 450.25,
450.49, 450.63 and 850.59.  Despite the balance of UV attenuation with
IR luminosity used in MAGPHYS to match photometry across all bands,
the IR luminosity in these systems is dramatically underestimated by
the best-fit \citet{bruzual03a} templates.  Of these sources, only
450.23 has an adequate photometric redshift estimate ($\Delta
z/(1+z)=0.03$), and it sits at $z<1$.  The other sources all have very
poor photometric redshift estimates ($\Delta z/(1+z)\approx0.60$) or
none at all.  While two of these sources only have far-IR
identifications at 450\um, and so could be false identifications as
DSFGs, three are robustly identified at both far-IR and
optical/near-IR wavelengths.  Thus, we attribute the cases of
catastrophic photometric redshift failures primarily to a geometric
decoupling of unobscured and obscured emission.

Included in Figure~\ref{fig:magphys} are far-infrared/millimeter SED
fits generated using a modified black body plus mid-infrared powerlaw,
as an independent assessment of the infrared luminosity, dust
temperature, and dust mass. We use the SED fitting procedure outlined
in \citet{casey12a}, using an emissivity spectral index of
$\beta=1.8$.  When there are fewer than 4 far-IR photometric points,
the mid-infrared powerlaw slope is fixed to $\alpha=2.5$ and when
there are fewer than 2 far-IR photometric points, the dust temperature
of the fit is fixed to $T_{\rm dust}=35$\,K.  A number of galaxies,
particularly those highlighted in the previous paragraph, are not well
fit at long wavelengths using the MAGPHYS energy balance approach but
are much better characterized at those wavelengths with this simple
SED approach.  We use these simple SEDs to characterize the IR
luminosities, dust temperatures, and dust masses of the sample, as
given in Table~\ref{tab:physical}.

\begin{table*}
\centering
\caption{Redshifts and Physically Derived Characteristics for MOSFIRE-confirmed DSFGs}
\begin{tabular}{l@{ }c@{ }c@{ }c@{ }c@{ }c@{ }c@{ }c@{ }c@{ }c}
\hline\hline
{\sc Source} & $z_{\rm spec}$ & {\sc Identifying Features} & M$_{\star}$ & M$_{\rm dust}$ & L$_{\rm IR}$ & SFR    & $\lambda_{\rm peak}$ & $\log({\rm IRX})$ \\
       &     &                            & [\msun]     & [\msun]        & [\lsun]      & [\sfr] & [\um]                &     \\
\hline 
450.05 & 2.551 & H$\alpha$, Ly$\alpha$ 
       &  (3.4$^{+0.1}_{-0.9}$)$\times10^{11}$ & (1.4$^{+3.7}_{-2.9}$)$\times10^{9}$ & (6.3$^{+2.4}_{-1.7}$)$\times10^{12}$ & 593$^{+226}_{-163}$ & 104$\pm$14 & 5.4$\pm$3.2 \\
450.06 & 2.092 & \ha,\nii,S{\sc II}, {\sc CIII]}, Fe{\sc II}, Mg{\sc II} 
       &  (1.1$^{+0.1}_{-0.2}$)$\times10^{11}$ &  (4.9$^{+0.8}_{-0.7}$)$\times10^{8}$ & (7.7$^{+1.2}_{-1.0}$)$\times10^{12}$ & 724$^{+112}_{-97}$ & 84$\pm$6 & 2.32$\pm$0.06 \\
450.09 & 2.473 & \ha,\hb 
       & (2.2$^{+0.1}_{-0.4}$)$\times10^{11}$ & (3.0$\pm$0.5)$\times10^{8}$ & (8.3$^{+2.6}_{-2.0}$)$\times10^{12}$ & 783$^{+243}_{-186}$ & 73$\pm$11 & 2.91$\pm$0.13 \\
450.14 & 1.523 & \ha,\nii 
       & $<$2.8$\times10^{11}$ & $<$2.7$\times10^{8}$ & $<$1.4$\times10^{11}$ & $<$128 & 108$\pm$43 & 2.30$\pm$0.38 \\
       & 2.462 & \ha,\nii 
       & ... & ... & ... & ... & ... & ... \\
450.21 & 0.628 & Pa$\beta$ 
       & (3.1$^{+0.4}_{-0.8}$)$\times10^{10}$ & (2.8$^{+5.4}_{-1.8}$)$\times10^{8}$ & (1.4$^{+1.1}_{-0.6}$)$\times10^{11}$ & 13$^{+10}_{-5}$ & 176$\pm$30 & 1.50$\pm$0.24 \\
450.23 & 0.9104 & \oii 
       & (5.7$^{+1.0}_{-1.0}$)$\times10^{8}$ & (3.2$^{+15.6}_{-2.7}$)$\times10^{8}$ & (7$^{+30}_{-5}$)$\times10^{11}$ & 63$^{+282}_{-52}$ & 182$\pm$38 & 2.76$\pm$0.74 \\
450.24 & 0.1658 & Pa$\alpha$, Pa$\beta$, H$_2$ 
       & (4.0$^{+0.6}_{-1.0}$)$\times10^{10}$ & (1.7$^{+0.6}_{-0.5}$)$\times10^{7}$ & (5.6$\pm$0.4)$\times10^{12}$ & 5.2$\pm$0.4 & 132$\pm$4 & 3.76$\pm$0.20 \\
450.25 & 1.516 & \ha, \nii, S{\sc II} 
       & (3.4$^{+0.1}_{-0.4}$)$\times10^{11}$ & (9.0$^{+2.0}_{-1.6}$)$\times10^{7}$ & (1.7$^{+0.9}_{-0.6}$)$\times10^{12}$ & 157$^{+80}_{-53}$ & 84$\pm$13 & 1.75$\pm$0.18\\
450.27 & 1.535 & \ha, \nii, S{\sc II}, Fe{\sc II}, Mg{\sc II} 
       & (3.0$^{+0.3}_{-0.6}$)$\times10^{11}$ & (4.5$^{+0.8}_{-0.7}$)$\times10^{8}$ & (4.0$\pm$0.5)$\times10^{12}$ & 373$^{+51}_{-45}$ & 97$\pm$6 & 2.39$\pm$0.06 \\
450.28 & 2.472 & \ha, \hb
       & (6.6$^{+0.3}_{-0.7}$)$\times10^{10}$ & (3.8$^{+2.9}_{-1.6}$)$\times10^{9}$ & (2.4$^{+1.9}_{-1.1}$)$\times10^{12}$ & 229$^{+176}_{-99}$ & 147$\pm$40 & 1.78$\pm$0.25 \\
450.37 & 2.109 & \ha,S{\sc II},\oiii,\hb 
       & (2.0$^{+0.5}_{-0.2}$)$\times10^{9}$ & (9.0$^{+4.9}_{-3.2}$)$\times10^{7}$ & (1.4$^{+8.4}_{-1.2}$)$\times10^{12}$ & 132$^{+787}_{-113}$ & 66$\pm$34 & 1.59$\pm$0.84 \\
450.49 & 1.554 & \ha 
       & (2.3$^{+0.1}_{-0.1}$)$\times10^{9}$ & (9.0$^{+4.9}_{-3.2}$)$\times10^{7}$ & (8$^{+63}_{-7}$)$\times10^{11}$ & 75$^{+593}_{-66}$ & 66$\pm$34 & 2.23$\pm$0.95 \\
450.58 & 2.464 & \ha 
       & (2.1$^{+0.1}_{-0.3}$)$\times10^{10}$ & (7.1$^{+2.8}_{-1.4}$)$\times10^{8}$ & (3.2$^{+1.5}_{-1.0}$)$\times10^{12}$ & 302$^{+145}_{-98}$ & 293$\pm$14 & $>$3.9 \\
450.59 & 0.4768 & \oii,\hb,\hb 
       & (2.3$^{+0.2}_{-0.8}$)$\times10^{10}$ & (6.8$^{+4.5}_{-2.7}$)$\times10^{7}$ & (1.1$^{+0.7}_{-0.4}$)$\times10^{11}$ & 11$^{+6}_{-4}$ & 105$\pm$19 & 1.66$\pm$0.20 \\
450.63 & 1.358 & \ha 
       & (4.8$^{+0.3}_{-1.0}$)$\times10^{8}$ & $<$9$\times10^{6}$ & (6$^{+57}_{-5}$)$\times10^{11}$ & 59$^{+541}_{-53}$ & $\equiv$90 & 1.91$\pm$1.00 \\
450.77 & 1.566 & \ha 
       & (4.9$^{+0.5}_{-1.2}$)$\times10^{10}$ & $<$9$\times10^{6}$ & (8$^{+89}_{-7}$)$\times10^{11}$ & 79$^{+836}_{-72}$ & $\equiv$90 & 2.85$\pm$1.07 \\
850.04 & 1.436 & \ha 
       & (3.0$^{+0.4}_{-0.7}$)$\times10^{10}$ & (2.8$^{+4.7}_{-1.7}$)$\times10^{9}$ & (1.1$\pm$0.2)$\times10^{12}$ & 106$^{+21}_{-17}$ & 167$\pm$10 & 2.00$\pm$0.08 \\ 
850.20 & 2.484 & \ha,\nii 
       & (3.2$^{+0.1}_{-0.8}$)$\times10^{10}$ & (3.2$^{+0.5}_{-0.4}$)$\times10^{8}$ & (1.2$\pm$0.2)$\times10^{12}$ & 1096$^{+166}_{-144}$ & 67$\pm$5 & 2.59$\pm$0.07 \\
850.25 & 2.438 & \ha 
       & (2.8$^{+0.1}_{-0.4}$)$\times10^{11}$ & (1.1$^{+0.3}_{-0.2}$)$\times10^{9}$ & (5.5$^{+2.2}_{-1.6}$)$\times10^{12}$ & 514$^{+211}_{-150}$ & 103$\pm$15 & 2.10$\pm$0.15 \\
850.36 & 0.224 & \ha,\oii 
       & (2.8$^{+0.1}_{-0.5}$)$\times10^{8}$ & (1.3$^{+3.2}_{-0.9}$)$\times10^{8}$ & (8.2$^{+9.7}_{-4.5}$)$\times10^{9}$ & 0.77$^{+0.92}_{-0.42}$ & 271$\pm$46 & 2.04$\pm$0.34 \\
850.44 & 2.470 & \ha,\nii,\oiii,\hb,H$\gamma$,\lya 
       & (1.8$^{+0.1}_{-0.4}$)$\times10^{10}$ & (2.0$^{+0.8}_{-0.6}$)$\times10^{8}$ & (2.8$^{+2.1}_{-1.2}$)$\times10^{12}$ & 263$^{+202}_{-114}$ & 91$\pm$26 & 1.41$\pm$0.25 \\
850.45 & 1.258 & \ha, \nii 
       & (4.7$^{+0.5}_{-1.0}$)$\times10^{10}$ & (4.3$^{+1.9}_{-1.3}$)$\times10^{8}$ & (5.4$^{+3.4}_{-2.1}$)$\times10^{11}$ & 51$^{+32}_{-20}$ & 140$\pm$25 & 2.22$\pm$0.22 \\
850.59 & 3.584 & \lya 
       & (5.8$^{+0.3}_{-1.1}$)$\times10^{10}$ & (5.1$^{+1.6}_{-1.2}$)$\times10^{8}$ & (3.5$^{+4.2}_{-1.9}$)$\times10^{12}$ & 333$^{+395}_{-181}$ & 92$\pm$40 & 1.49$\pm$0.34 \\
850.60 & 1.457 & \ha,\nii, S{\sc ii} 
       & (2.1$^{+0.1}_{-0.2}$)$\times10^{11}$ & (2.4$\pm$0.4)$\times10^{8}$ & (2.3$^{+0.4}_{-0.4}$)$\times10^{11}$ & 218$^{+42}_{-35}$ & 96$\pm$8 & 2.96$\pm$0.09 \\
850.82 & 2.508 & \ha 
       & (1.8$^{+0.2}_{-0.3}$)$\times10^{11}$ & (2.8$^{+0.7}_{-0.6}$)$\times10^{8}$ & (5.0$^{+3.4}_{-2.0}$)$\times10^{12}$ & 471$^{+323}_{-191}$ & 80$\pm$24 & 2.71$\pm$0.23 \\
850.89 & 2.104 & \ha,\nii,\oiii,\hb 
       & (1.8$^{+0.1}_{-0.4}$)$\times10^{11}$ & (6.8$^{+2.0}_{-1.5}$)$\times10^{8}$ & (3.7$^{+1.8}_{-1.2}$)$\times10^{12}$ & 347$^{+173}_{-115}$ & 104$\pm$19 & 1.54$\pm$0.18 \\
850.95 & 1.556 & \ha, \nii, NaD 
       & (3.8$^{+0.4}_{-1.0}$)$\times10^{10}$ & (2.8$^{+1.0}_{-0.7}$)$\times10^{8}$ & (2.0$^{+1.4}_{-0.8}$)$\times10^{12}$ & 189$^{+131}_{-77}$ & 102$\pm$28 & 1.66$\pm$0.29 \\
850.97 &  0.731  & \oii
       & (1.5$^{+0.1}_{-0.3}$)$\times10^{11}$ & (4.7$^{+6.4}_{-4.3}$)$\times10^{8}$ & (2.8$^{+29.7}_{-2.5}$)$\times10^{11}$ & 5.8$^{+26.1}_{-4.7}$ & 109$\pm$15 & 2.9$\pm$1.1 \\
m450.87 & 2.282 & \ha, \oiii 
       & (2.0$^{+0.4}_{-0.8}$)$\times10^{11}$ & (6.5$^{+1.4}_{-1.2}$)$\times10^{8}$ & (5.4$^{+1.7}_{-1.3}$)$\times10^{12}$ & 512$^{+161}_{-122}$ & 94$\pm$13 & 3.21$\pm$0.18 \\
m450.133 & 0.728 & Pa$\beta$,\hb,\oiii,\oii,Mg{\sc II} 
       & (6.3$^{+5.5}_{-1.1}$)$\times10^{10}$ & (7.5$^{+2.7}_{-2.0}$)$\times10^{7}$ & (2.5$^{+1.5}_{-0.9}$)$\times10^{11}$ & 23$^{+14}_{-9}$ & 127$\pm$24 & 0.73$\pm$0.21 \\ 
m450.173 & 1.003 & \oii, NeV, NeIII, \hd, Mg{\sc II} 
       & (2.2$^{+0.3}_{-0.4}$)$\times10^{10}$ & (1.6$^{+0.7}_{-0.5}$)$\times10^{8}$ & (6.5$^{+5.2}_{-2.9}$)$\times10^{11}$ & 61$^{+49}_{-27}$ & 116$\pm$20 & 1.38$\pm$0.26 \\
\hline\hline
\end{tabular}
\label{tab:physical}

{\small {\bf Table Notes} --- Source names are as in \citet{casey13a},
  and the positions of the sources are given in their Tables~6-7; in
  all of the cases above, the most likely multiwavelength counterpart
  was chosen (counterpart 1/N).
The redshift identification is given to three significant digits
unless the source was particularly low redshift with bright emission
features which allow for a more precise redshift constraint.  
The `Identifying Features' column indicates which spectral features
were used as the primary lines in identifying the sources'
spectroscopic redshift.
Stellar mass estimates are generated with the help of {\sc Magphys}
energy balance code as described in \S~\ref{sec:photz}.
M$_{\rm dust}$, L$_{\rm IR}$, SFR and $\lambda_{\rm peak}$ are all
derived using a simple modified blackbody plus mid-infrared powerlaw
\citep{casey12a}, as shown in Figure~\ref{fig:magphys}.  We quote
$\lambda_{\rm peak}$ instead of dust temperature $T_{\rm dust}$ as the
former is more directly tied to observational constraints and the
latter is heavily dependent on model assumptions.  The SFR quoted here
is derived from $L_{\rm IR}$ using a Chabrier IMF \citep{kennicutt12a}.
The last column, $\log({\rm IRX})$, is the log of the IRX ratio,
defined as $L_{\rm IR}$/$L_{\rm UV}$.  For example, a DSFG with
$\log(IRX)=2$ will have an IR luminosity 100 times its UV luminosity.
}
\end{table*}

\section{Characteristics of the Spectroscopic Sample}\label{sec:individual}

In this section we describe some of the unique characteristics of the
spectroscopically-confirmed COSMOS DSFGs listed in
Table~\ref{tab:physical}.  The sample is naturally heterogeneous, and
thus requires some individual and subset remarks to fully capture the
physical nature of their evolution and the source of their obscured
emission.  We group sources by common characteristics, and include
some remarks on ambiguous and mis-identifications at the end of this
section.

\subsection{Unlensed DSFGs above $L>10^{11}$\lsun}\label{sec:dsfgs}

\begin{figure}
\centering
\includegraphics[width=0.99\columnwidth]{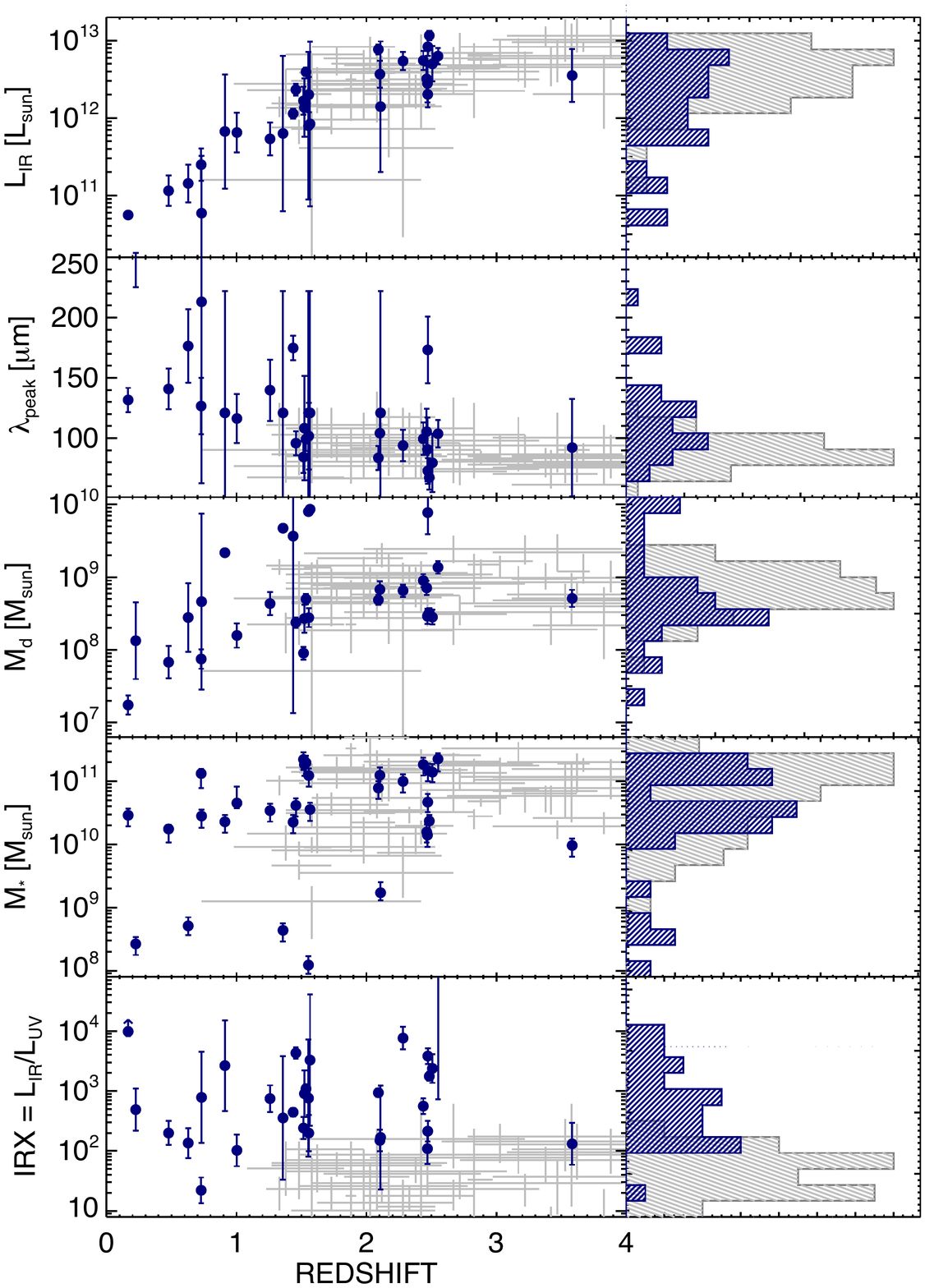}
\caption{Basic measured physical characteristics of the
  spectroscopically-confirmed DSFG sample from this paper (navy)
  against a comparison sample of photometric sources from the ALESS
  survey \citep{da-cunha15a}.  At top, total IR luminosity ranges from
  a few times $10^{10}$\,\lsun\ to 10$^{13}$\,\lsun.  The peak
  rest-frame SED wavelength falls towards high-redshifts; the
  450\um\ flux density points are crucial to the measurement of this
  quantity.  Dust masses average several times $10^{8}$\,\msun, about
  0.1\%\ of the galaxies' stellar masses; compared to the ALESS
  sample, this SCUBA-2 sample seems slightly less massive in dust for
  comparable stellar masses.  At bottom, IRX, or the ratio of $L_{\rm
    IR}/L_{\rm UV}$, is quite high to the ALESS sample, converted from
  $A_{V}$ to $IRX$.}
\label{fig:lirz}
\end{figure}

DSFGs that have intrinsic infrared luminosities exceeding
$L>10^{11}$\lsun\ can be characterized as bona-fide DSFGs: luminous
dusty galaxies, predominantly at a redshift beyond $z>1$. Twenty-nine
galaxies in our sample fall into this luminosity regime and range in
redshift from 450.24 at $z=0.166$ to 850.89 at $z=3.584$, with a
median redshift of $\langle z\rangle = 1.6$.  Their luminosities, dust
temperature constraints, dust masses, and $L_{\rm IR}/L_{\rm UV}$
ratios are shown in Figure~\ref{fig:lirz}.  The median peak of the
rest-frame far-infrared SED is 116$\pm$9\um, which translates to
$\sim$36\,K dust temperature using our model assumptions.  On average,
the sample is 130$^{+40}_{-30}$ times more luminous in the
far-infrared/submillimeter than in the UV/optical, consistent with
most luminous DSFGs. Figure~\ref{fig:lirz} also compares
  our spectroscopic sample with the ALESS photometric sample of
  \citet{da-cunha15a}; our sample sits at lower redshifts, and thus
  reaches to slightly lower luminosities, slightly cooler dust
  temperatures, lower dust masses, and yet higher obscurations of
  their rest-frame UV emission.

\begin{figure*}
\centering
\includegraphics[width=0.59\columnwidth]{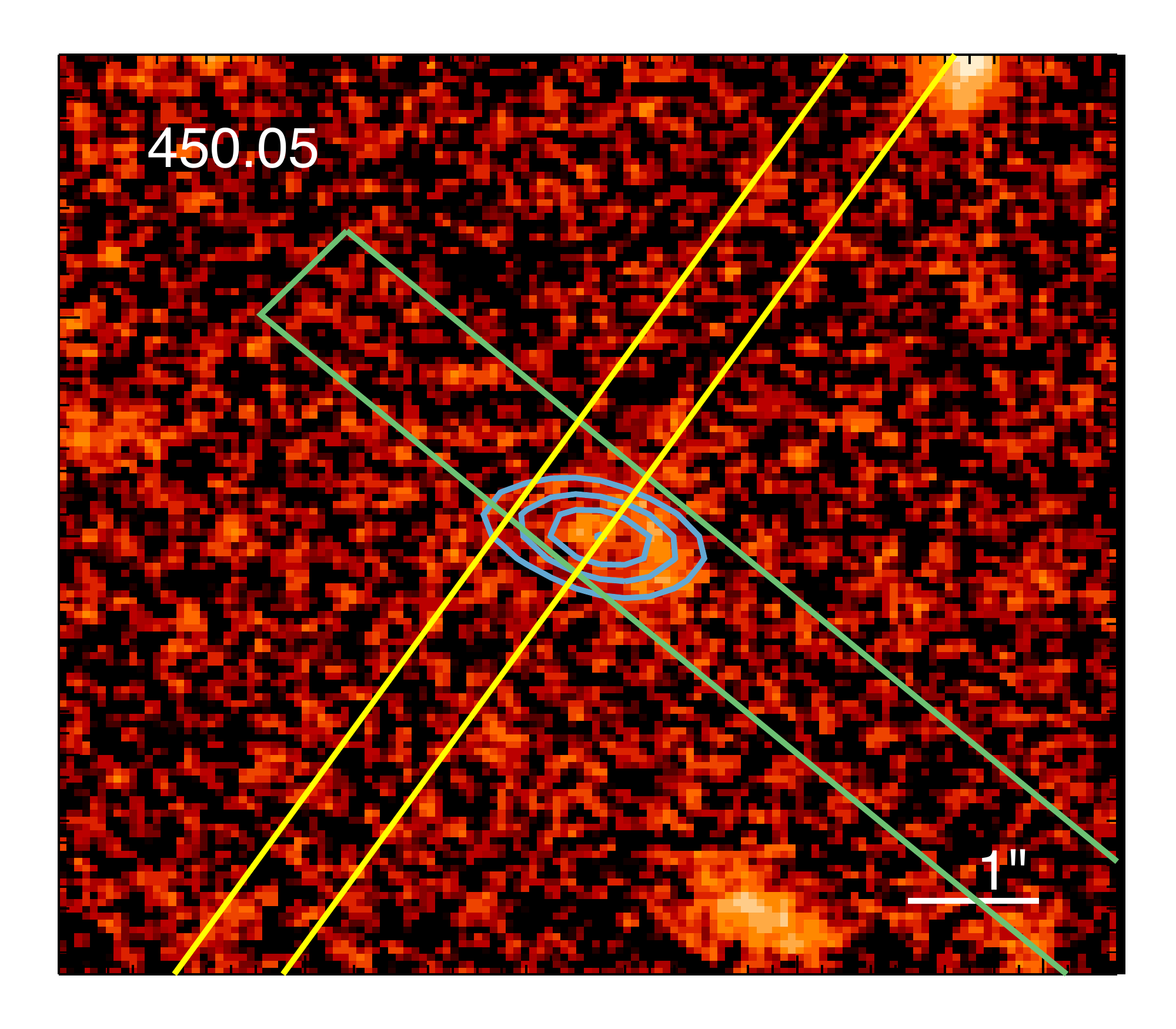}
\includegraphics[width=0.59\columnwidth]{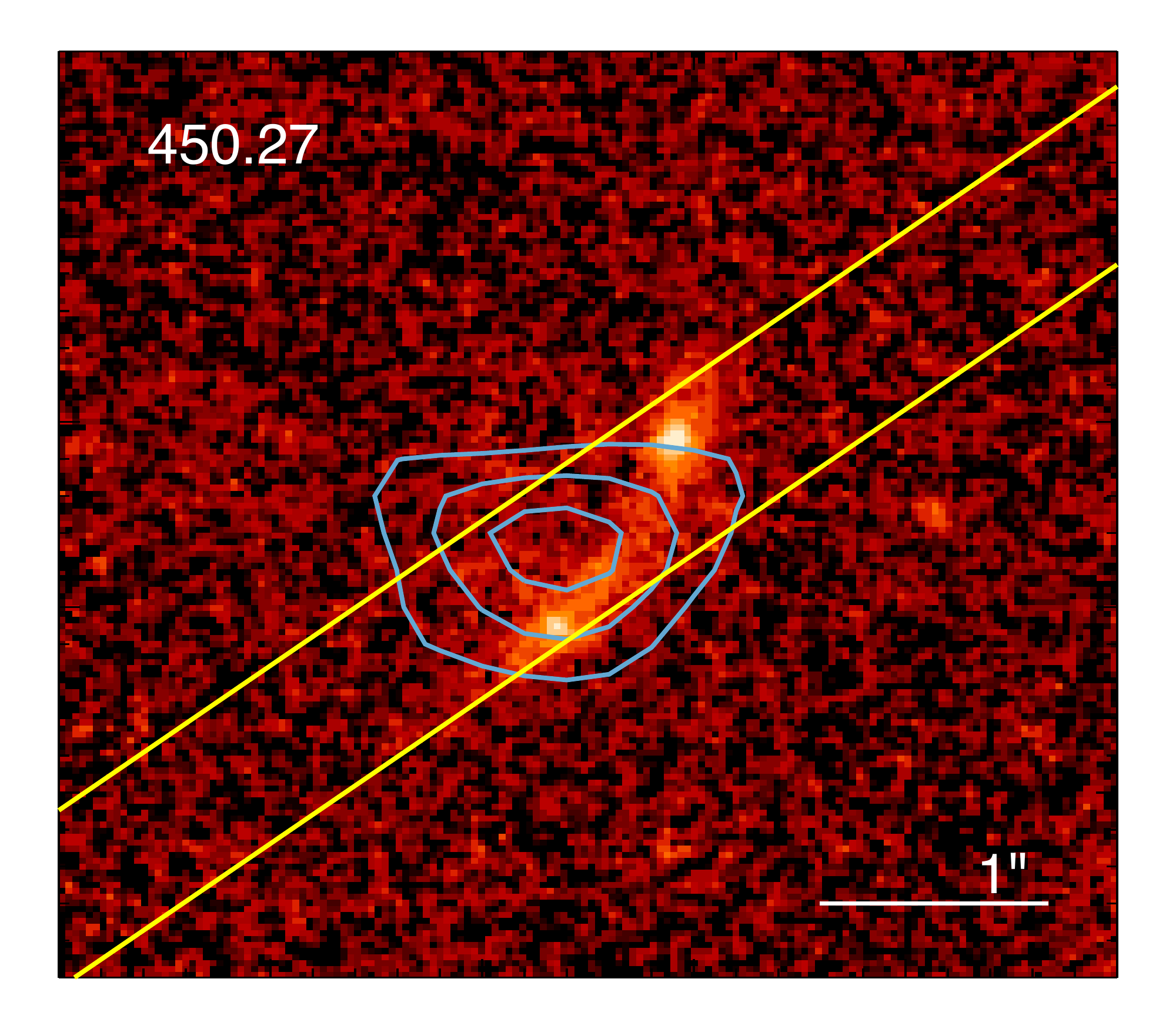}
\includegraphics[width=0.59\columnwidth]{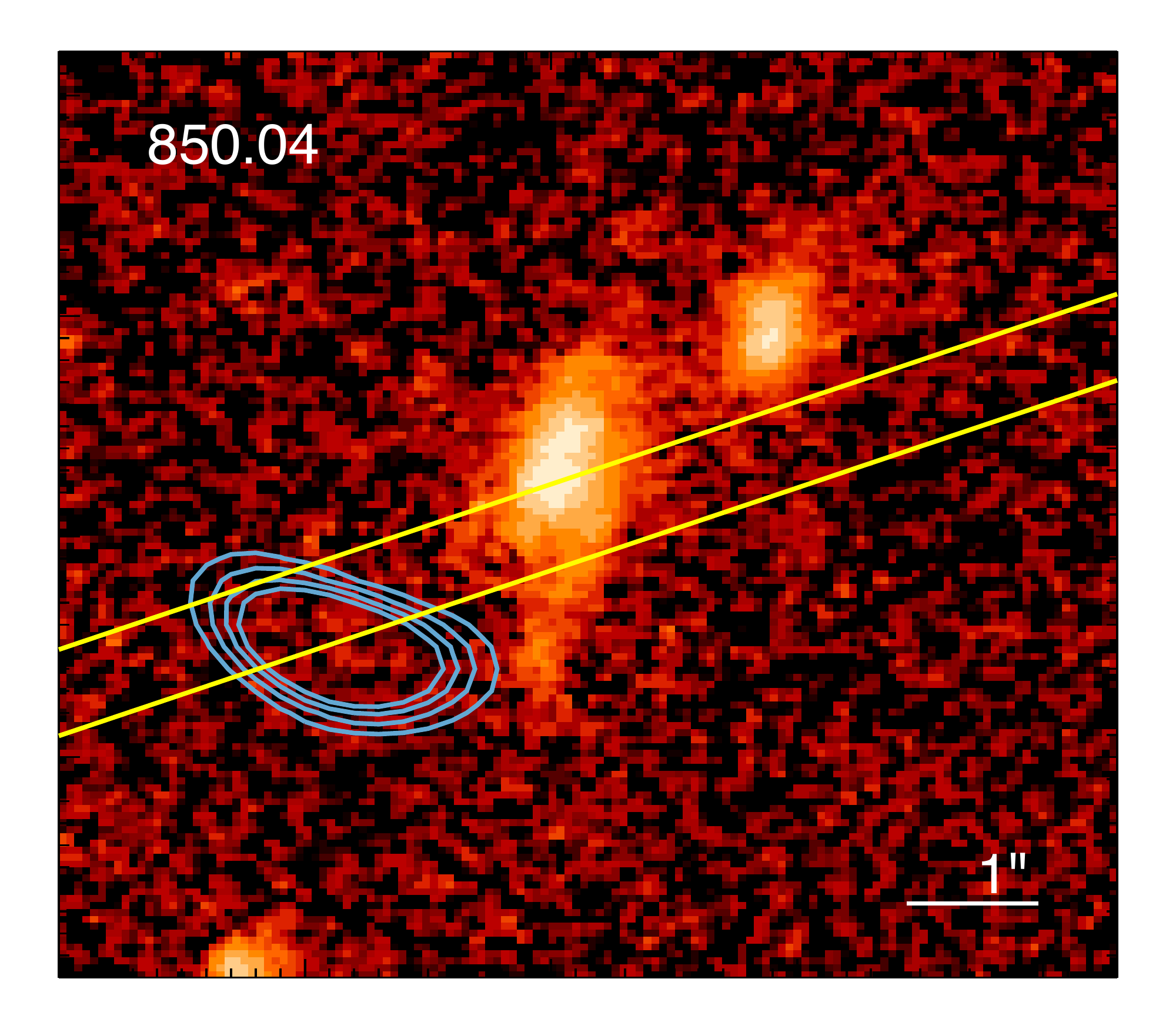}\\
\includegraphics[width=0.59\columnwidth]{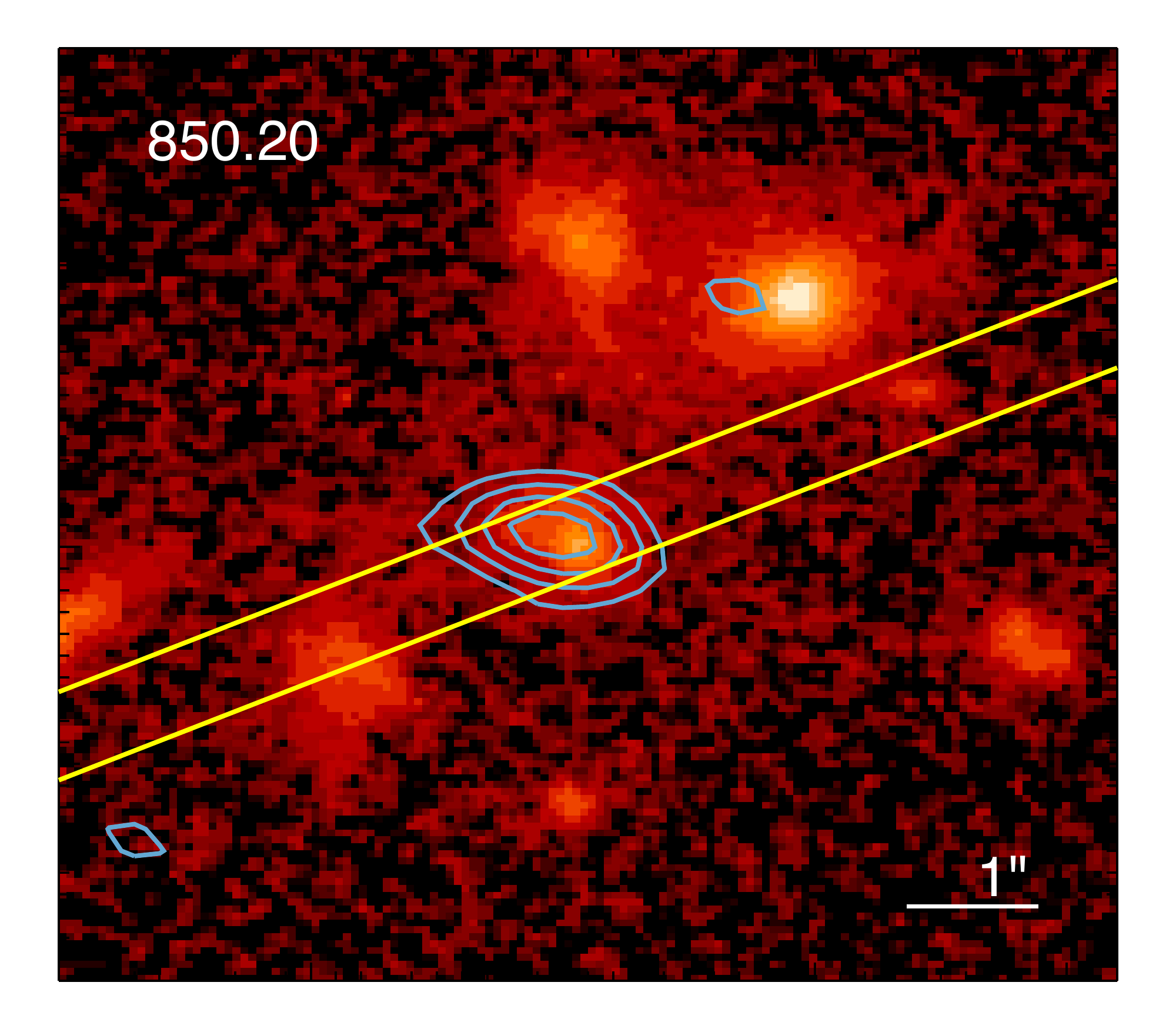}
\includegraphics[width=0.59\columnwidth]{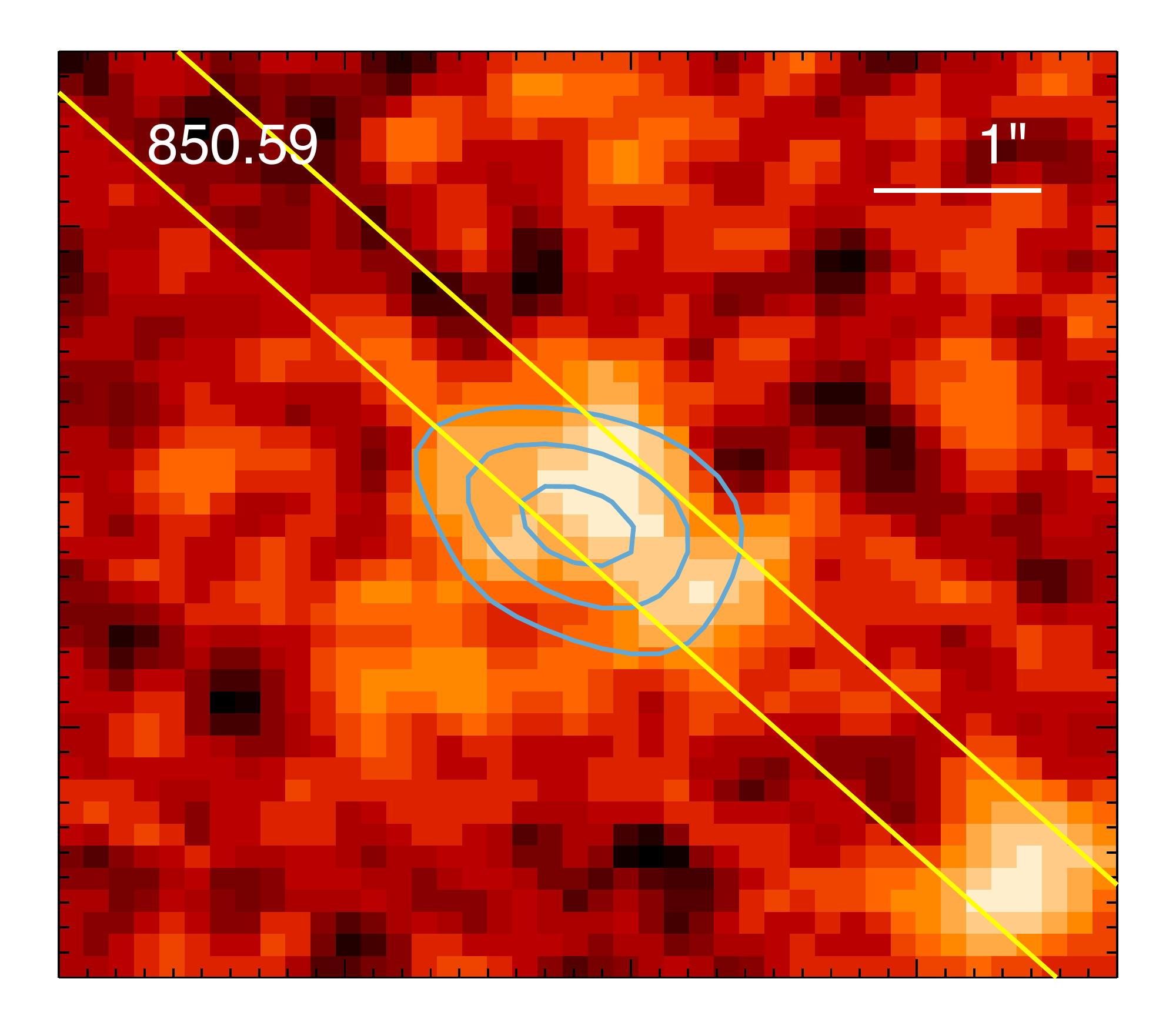}
\caption{ Sources from our spectroscopic sample with ALMA
  observations, confirming their positions to align with those
  identified via multiwavelength prior techniques
  \citep[in][]{casey13a}. Cutouts range in size from 4$''$ to 7$''$ to
  a side, as indicated by the inset arcsecond marker.  The first four
  cutouts are from HST F125W imaging from the CANDELS survey, and the
  last cutout is from ground-based UltraVISTA $H$-band imaging.  ALMA
  band 6 (1.1\,mm) dust continuum contours are shown in blue, MOSFIRE
  slits in yellow, and LRIS slits in green. From left to right, source
  450.05, source 450.27, source 850.04, source 850.20, and source
  850.59.  While the submillimeter emission in source 850.04 is offset
  from its OIR counterpart, we present a further morphological and
  kinematic analysis of this source in Drew \etal, in preparation,
  that lead us to tentatively conclude the two components are
  associated.}
\label{fig:alma}
\end{figure*}

The median dust mass is (4.4$\pm$1.0)$\times$10$^{8}$\,\msun, which is
about 0.9$\pm$0.4\%\ of the median stellar mass,
(4.9$\pm$2.1)$\times10^{10}$\,\msun.  Assuming an average gas-to-dust
ratio of 100 \citep{scoville14a}, this would imply molecular gas
reservoirs averaging 4$\times10^{10}$\,\msun. With a median
star-formation rate of 220$\pm$60\,\sfr, the inferred representative
gas depletion time for this population is $\approx$180\,Myr, though we
stress the importance of measuring this directly with emission from
cold molecular gas.  Five sources in this category also have archival
ALMA 1.1\,mm dust continuum data pinpointing their submillimeter
emission to the positions of our MOSFIRE follow-up and are shown in
Figure~\ref{fig:alma}.  Two additional sources have ALMA 1.1\,mm
imaging which reveal misidentifications, as discussed in
\S~\ref{sec:misids}.

%
%


\subsection{Low-redshift DSFGs}

Two sources identified in our sample sit at very low redshifts
$z<0.3$.  850.36 has a measured spectroscopic redshift of $z=0.224$
from the zCOSMOS survey \citep{lilly07a}; its identification as the
origin of the submillimeter emission is unambiguous.  The optical
morphology is clearly consistent with a major merger with double
nuclei and an extended, asymmetrical disk.  While we did target this
source with MOSFIRE on the same mask as other higher priority DSFGs,
the wavelength coverage failed to overlap with expected prominent
features like Pa$\alpha$ and Pa$\beta$.

Source 450.24 also has a prior spectroscopic confirmation at
$z=0.1661$ \citep{lilly07a}, and here we obtained a near-infrared
spectrum which is more consistent with a redshift of $z=0.1657$ based
on detection of Pa$\alpha$, Pa$\beta$ and H$_2$ features.  We measure
a Paschen decrement
between Pa$\alpha$ and Pa$\beta$ of 8.7$\pm$2.5,
with the significant uncertainty caused by flux calibration between bands,
and the Pa$\beta$ detection near the edge of $H$-band close to telluric
features.  While highly uncertain, this ratio is still significantly
higher than the expected theoretical intensity ratio of 2.16 for case
B recombination temperature of 10$^4$\,K, suggesting significant dust
obscuration.

Source 450.21 has a strong emission line detection at 2.087\um, though
this is not in agreement with a previously reported spectroscopic
redshift of $z=0.837$ (John Silverman, private communication).
We assign this source a tentative redshift of $z=0.628$.

\subsection{Spatially Superimposed}

\begin{figure}
\centering
\includegraphics[width=0.99\columnwidth]{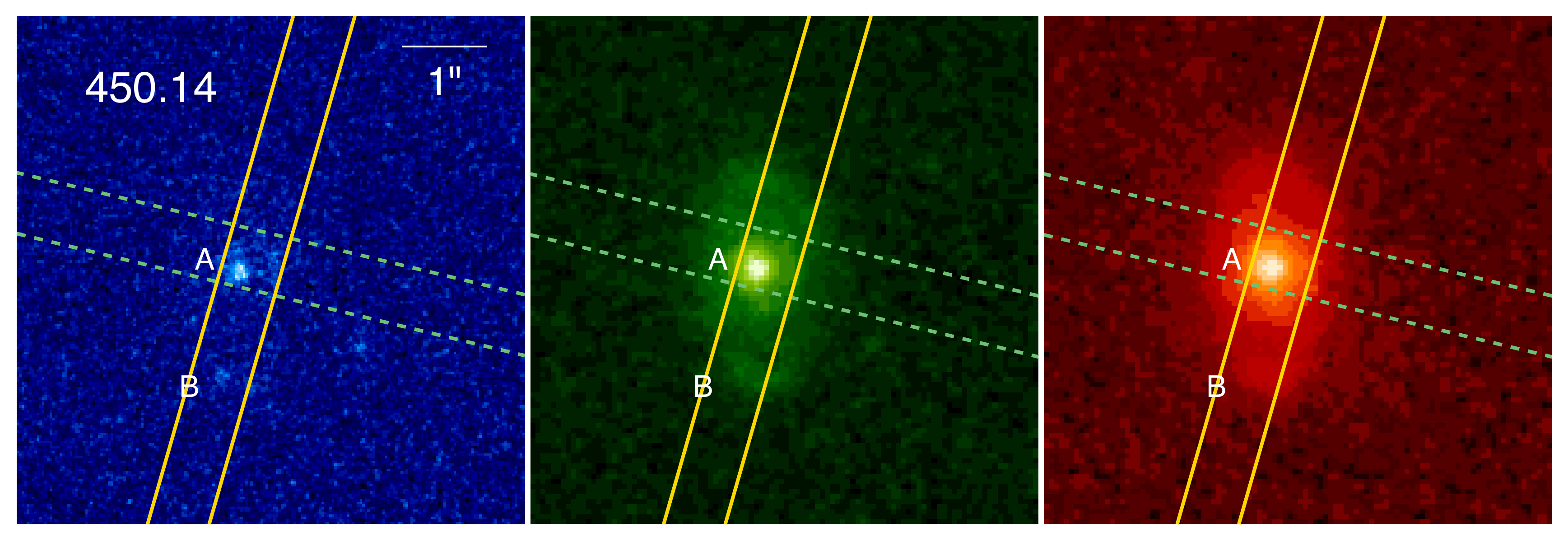}
\includegraphics[width=0.99\columnwidth]{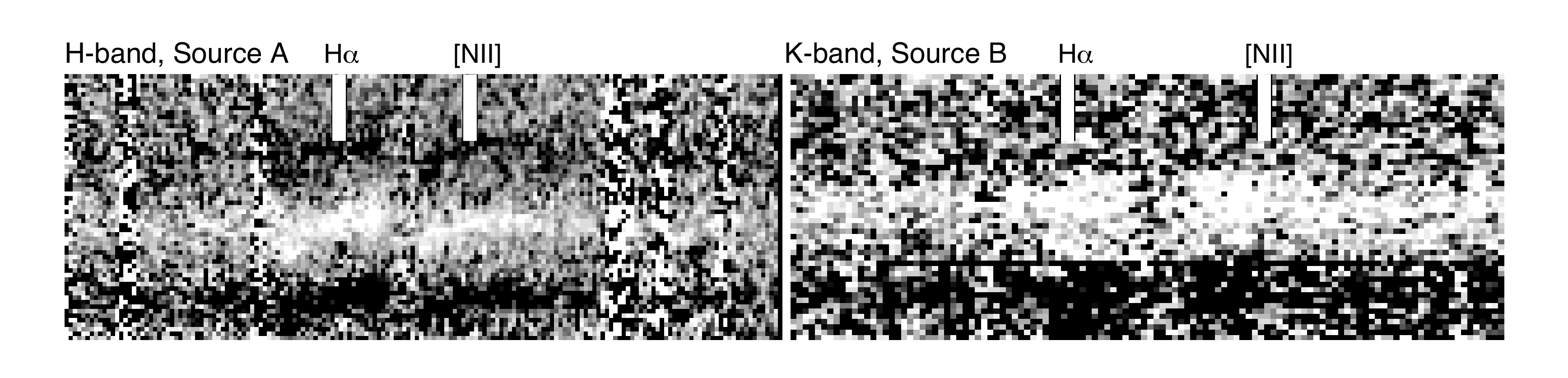}
\caption{Top: {\it HST} ACS and WFC3 cutouts around 450.14, an
  ambiguous case of a potentially superimposed pair of sources at
  $z=1.523$ and $z=2.462$.  Source A is identified unequivocally at
  $z=1.523$ via detection of \ha\ and \nii, shown in the lower panel.
  Source B is tentatively at $z=2.462$.  We overplot two different
  MOSFIRE slits we used to observe this source (though vignetting
  prevents us from comparing the $K$-band spectra of the two). Bottom:
  Spectra around \ha\ emission for Source A (left) and Source B
  (right).  \ha\ and \nii\ emission are marked at the measured
  redshifts.}
\label{fig:450-14}
\end{figure}

\begin{figure*}
\centering
\includegraphics[width=0.49\columnwidth]{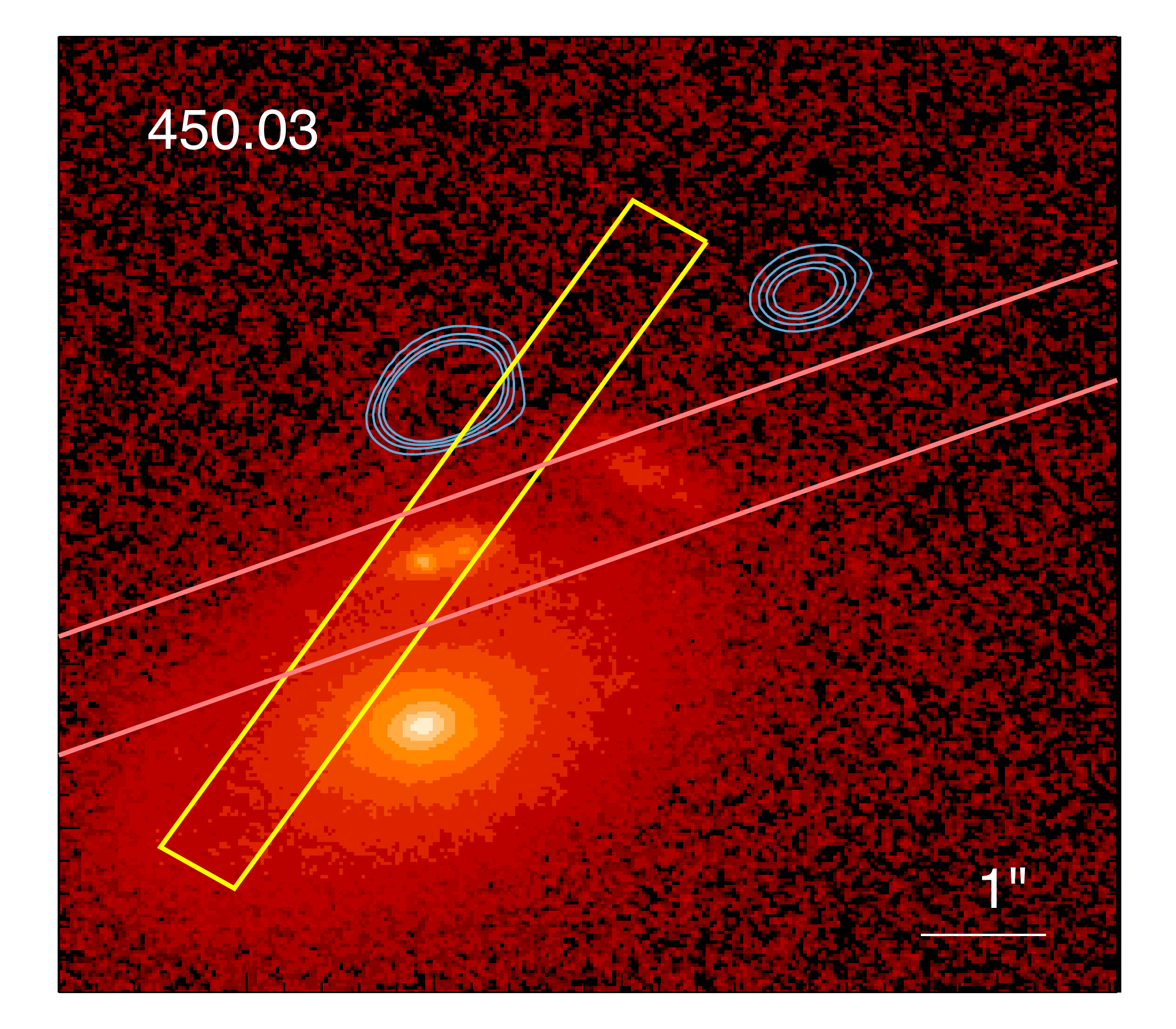}
\includegraphics[width=0.49\columnwidth]{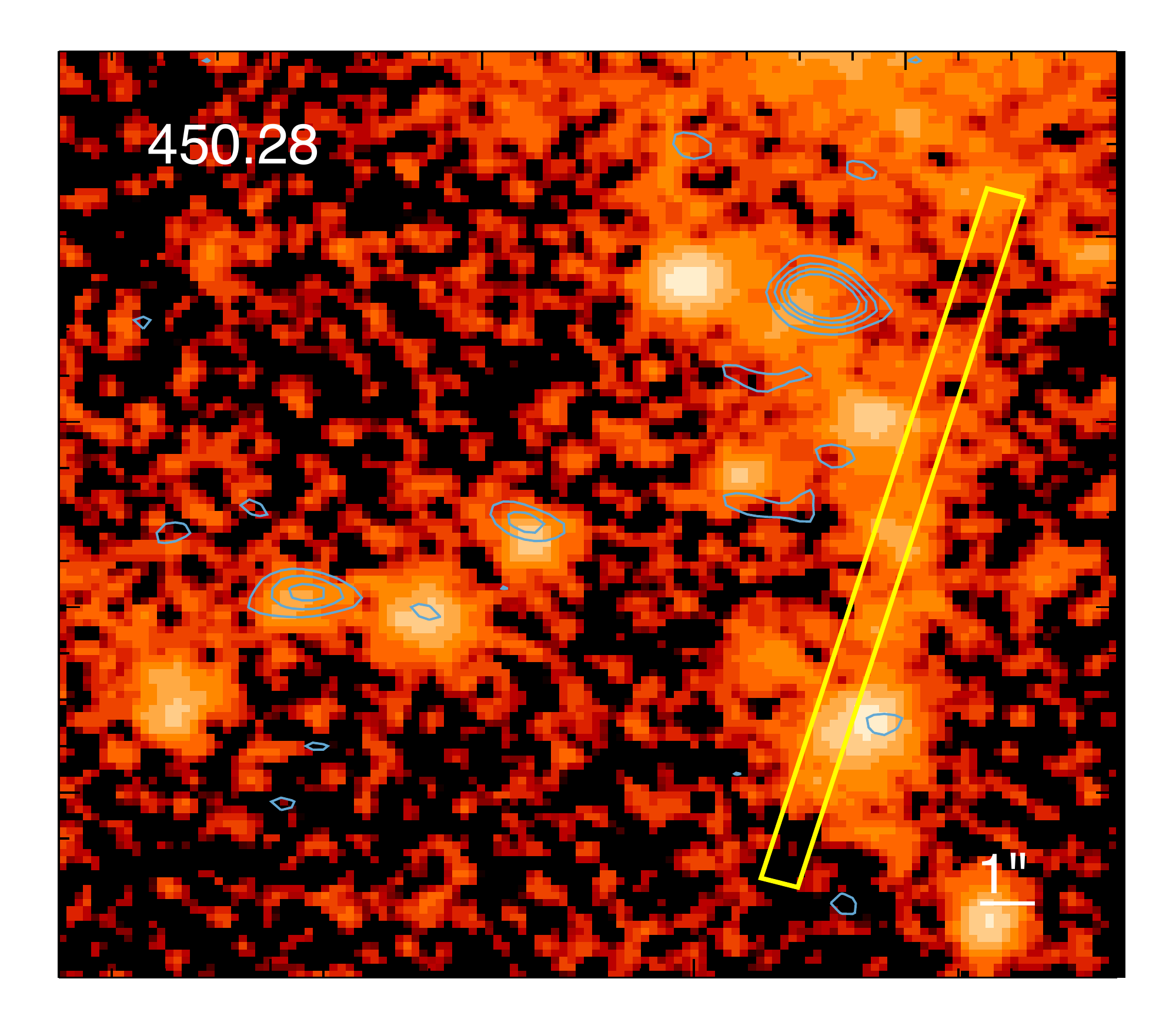}
\includegraphics[width=0.49\columnwidth]{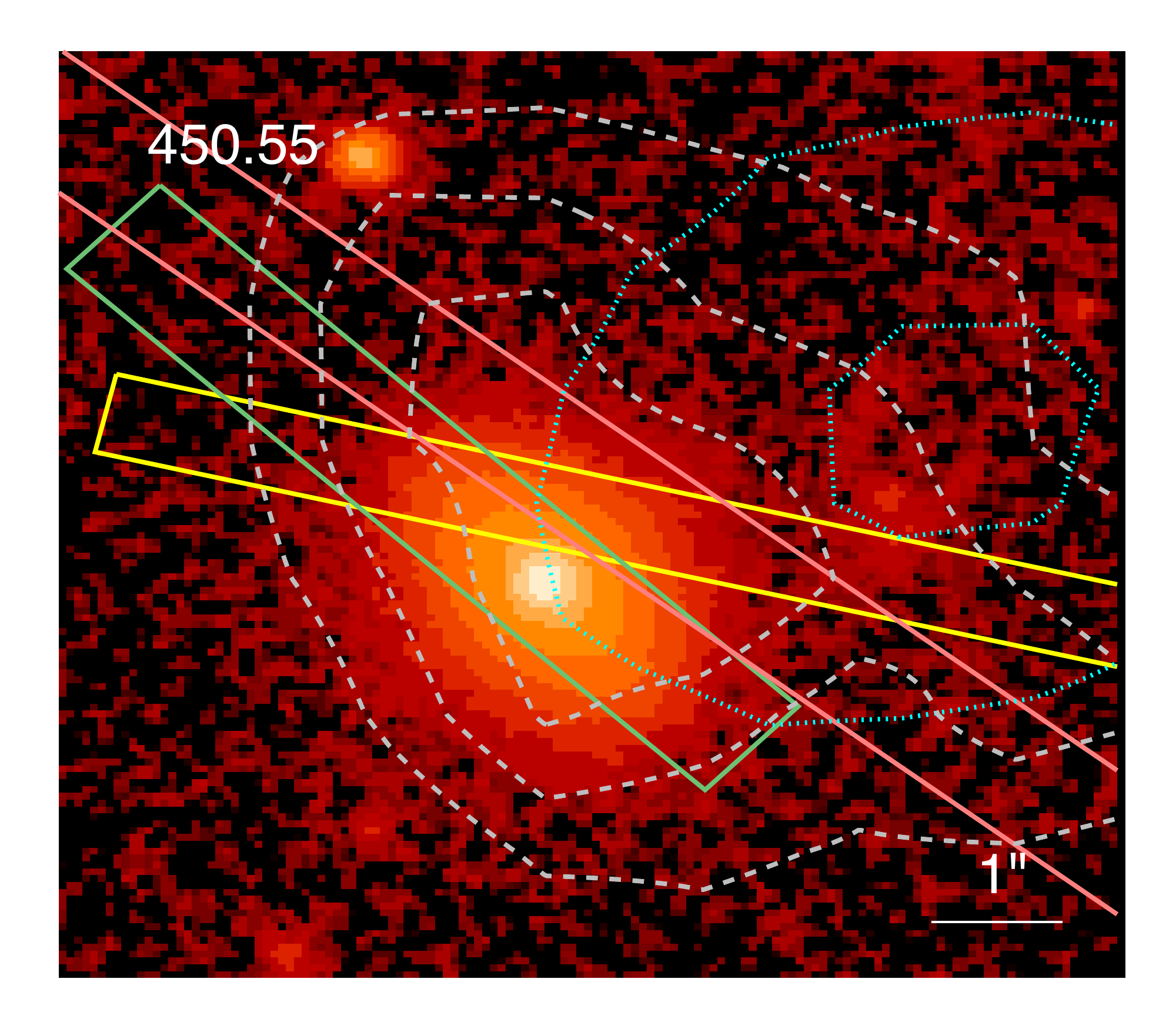}
\includegraphics[width=0.49\columnwidth]{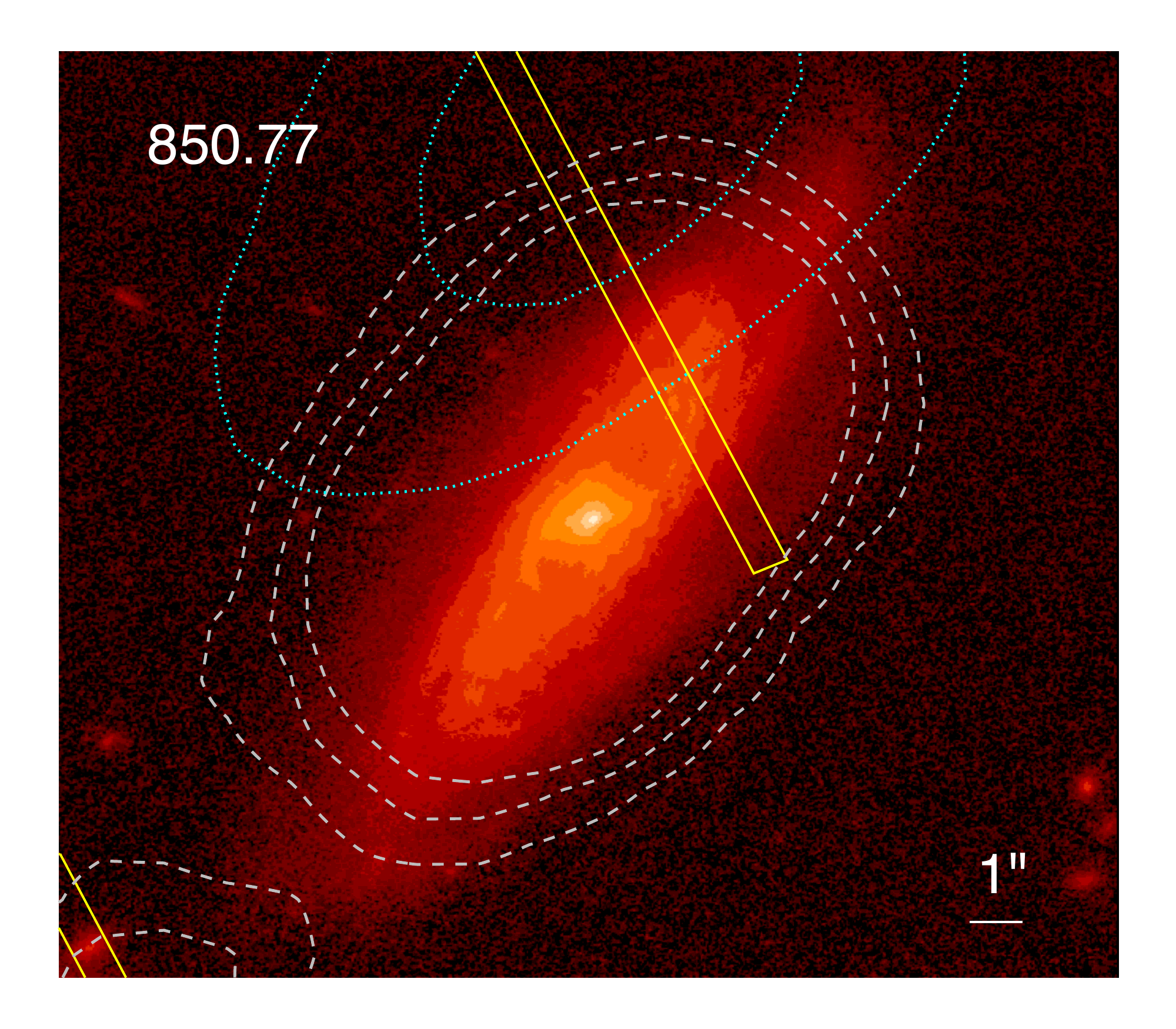}
\caption{A characterization of four mis-identifications. {\it Left:}
  Source 450.03, known as AzTEC-2 in the literature, was initially
  thought to be associated with the source at the center of the frame,
  aligned in both the MOSFIRE (yellow) and DEIMOS (pink) slit, and
  confirmed to be at $z=1.123$.  Long wavelength interferometric
  follow-up suggested the source sat further to the north, though
  these positions were thought to be inconsistent with the positional
  constraint at 450\um.  However, further higher-frequency ALMA
  follow-up (blue contours) reveal that indeed, the DSFG sits further
  to the north.  The background image is WFC3 1.25\um\ imaging from
  CANDELS.  
{\it Middle left:} Source 450.28 is part of a complex region of submm
blends, now revealed by ALMA to consist of multiple sources.
Unfortunately the source targeted in our MOSFIRE observations,
confirmed at $z=2.472$, is not properly identified as one of the submm
sources.  However, the brightest of the submillimeter sources has been
spectroscopically confirmed at $z=2.494$ \citep{wang16a}.
{\it Middle Right:} Source 450.55 was matched to a fairly bright
  $z=0.36$ galaxy which dominates the 24\um\ emission (dashed gray
  contours) of the area.  This counterpart is not very well aligned
  with the submillimeter emission (dotted cyan contours) and is likely
  to be a foreground contaminant.  Unfortunately our spectroscopic
  program failed to account for this foreground source before carrying
  out observations.
{\it Right:} Source 850.77 is a submillimeter source without an
obvious multiwavelength counterpart, centered on the outskirts of a
nearby spiral galaxy. }
\label{fig:misids}
\end{figure*}

One DSFG in our sample is an ambiguous potential overlap of two
sources at different redshifts.  The optical/near-infrared counterpart
to the submillimeter source is shown in Figure~\ref{fig:450-14}; it
consists of a central compact source with two fainter arcs to the
north and south. We have unequivocally identified the central compact
source at a redshift of $z=1.523$.  The arcs are much more visible in
the WFC3 imaging of the source, suggesting a higher redshift solution
might be more plausible.  This target was placed on two different
MOSFIRE masks, and one happened to align with the southern faint arc.
We tentatively suggest that this fainter arc is a background source
sitting at $z=2.462$ as identified by two anomalous emission features
in the $K$-band spectrum consistent with \ha\ and \nii\ at this
alternate redshift. Figure~\ref{fig:450-14} also shows the sources' 2D
spectrum, zoomed in to the H$\alpha$ features in $H$- and $K$-bands for
the respective redshifts of the central source and potential
background source.  Though the detection of H$\alpha$ and \nii\ are
only tentative for the $z=2.462$ identification, there is also a
tentative corresponding detection of \oiii\ in $H$-band at the same
redshift.  This source will require more substantial follow-up to
determine if the spatial superposition is genuine.  Other similar
spatially overlapping sources have been found in submillimeter
redshift surveys \citep[e.g. the ALESS sample;][]{danielson16a}.

\subsection{Mis-identifications}\label{sec:misids}

Inevitably the positional uncertainty in bolometer maps will lead to
some mis-identifications and ambiguities which propagate to
spectroscopic follow-up campaigns.  Here we present two
mis-identifications, along with two ambiguous cases that are likely
mis-identifications.  All are shown in 
Figure~\ref{fig:misids}.

Source 450.03 (Figure~\ref{fig:misids}, left) is best known as AzTEC-2
in the literature \citep{younger07a,younger09a}, and has extensive
follow-up, both interferometric and spectroscopic.  Unfortunately
spectroscopic efforts to-date have failed to yield a redshift
identification.  At the time of our MOSFIRE and DEIMOS campaign, it
seemed like the interferometric source (at 1.1mm) was inconsistent
with the 450\um\ position identified in \citet{casey13a}, so the
possibility remained that they were separate sources.  More recent
high-frequency follow-up from ALMA at 850\um\ indicates this is not
the case, however, and the position of the DSFG sits about
1-1.5\arcsec\ to the north of our spectroscopic target.  While we have
confirmed our original target sits at $z=1.123$, and there is a
possible detection of a CO line at the same redshift (Jimenez \etal,
in preparation) our confirmation does not spatially align with the
DSFG, and so we do not include it in further analysis for this paper
or sample. The possibility remains this DSFG is
  associated with the system at $z=1.123$ but requires further
  analysis.

Source 450.28 (Figure~\ref{fig:misids}, middle left) is a member of a
complex blend of submm sources; the prior measurements of this area at
850\um\ and 1.2\,mm only reveal a single source
\citep[see][]{bertoldi07a,casey13a}, while 450\um\ imaging breaks the
emission into two distinct peaks.  Our MOSFIRE follow-up targeted the
most likely counterpart for the western source, 450.28, and confirms a
redshift for that source of 2.472 \citep[a member of the protocluster
  discussed in][]{casey15a}.  However, more recent ALMA follow-up
reveals this source is mis-identified, and instead, three sources
dominate the ALMA emission to the north and east.  While formally
resulting in a mis-identification, this complex is physically
associated with the structure identified at $z\approx2.47$, confirmed
through multiple transitions of CO at $z=2.494$.  This region is
discussed at length in \citet{wang16a} as a potential virialized
protocluster core at $z=2.51$, where they present an extended X-ray
detection and eleven possible CO(1-0) detections in the broader
redshift range of the protocluster.  Further CO(1-0) observations of
this region is being analyzed to determine redshifts and further map
this complex region (Champagne \etal, in preparation).

Source 450.55 (Figure~\ref{fig:misids}, middle right) is likely a
blend of a background galaxy which is exceedingly faint, even at
near-infrared wavelengths, and a foreground galaxy, confirmed at
$z=0.36$.  The submillimeter source is more consistent with the
background source to the west of the bright foreground source, yet the
foreground source is 24\um-bright, and so was earlier identified as
the likely submillimeter counterpart.  Deblending of this source was
unfortunately not attempted before the spectroscopic campaign was
underway, so we have not obtained a secure redshift for this DSFG.

The last case of a mis-identification is 850.77.  This submillimeter
source is offset to the north of a very bright foreground spiral
galaxy at $z=0.1088$.  The spiral is clearly the dominant source of
24\um\ emission in the region, and so was mistaken as a potential
counterpart to the 850\um\ source.  Deep imaging does not reveal any
obvious optical/near-infrared counterpart, and so this source remains
ambiguous and without a redshift identification; it is also not
detected at 450\um.

\begin{figure*}
\includegraphics[width=1.99\columnwidth]{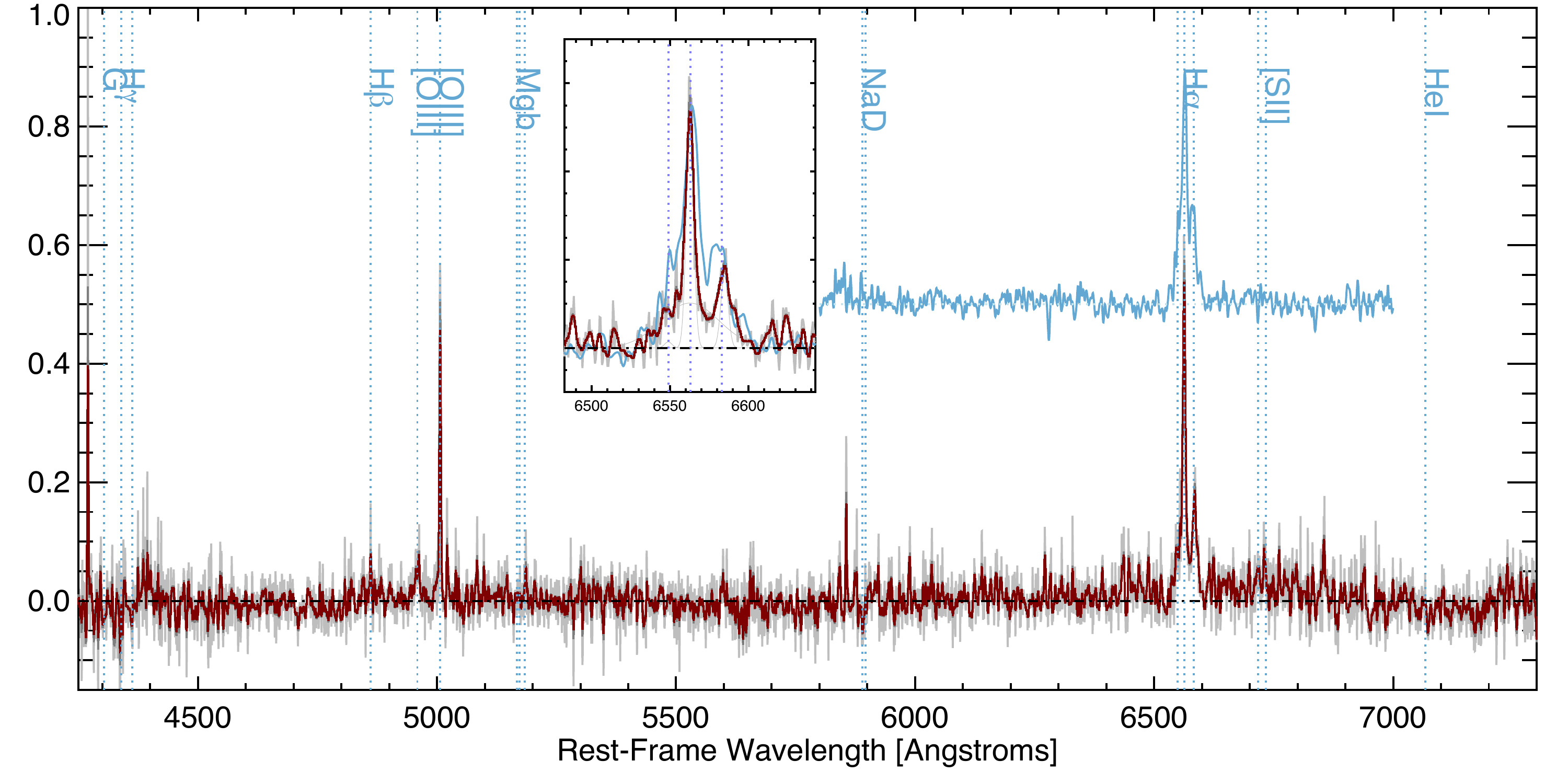}
\caption{The composite optical DSFG spectrum, comprised
  of twenty DSFGs within the redshift range $1.26<z<2.55$.  The median
  redshift of the stack is $\langle z\rangle=2.1$.  The composite is
  Gaussian smoothed to a spectral resolution of $\sim$1200 (dark red)
  from the original R$\sim$3600 (gray).  There are significant
  detections of \ha, \hb, \nii, \oiii, and \sii\ in emission, and Na\,I
  in absorption.  We compare our composite to that of
  \citet{swinbank04a}, shown in blue offset in flux.  A zoom-in of the
  region surrounding \ha\ and \nii\ emission is shown inset.  See the
  text, \S~\ref{sec:stack}, for more details.  This spectrum is available for download at {\tt www.as.utexas.edu/$\sim$cmcasey/downloads.html}.}
\label{fig:stackha}
\end{figure*}
\section{DSFG Composite Spectrum}\label{sec:stack}

With many rest-frame optical spectra of DSFGs in hand, stacking
provides a useful means of inferring the aggregate spectral
characteristics of the DSFG population, including features which are
too weak to measure in individual DSFG spectra.  We construct a
rest-frame optical spectrum for the 20 DSFGs with \ha\ detections by
first shifting all spectra into the rest-frame using the systemic
redshift measured from \ha\ emission.  The redshifts for the galaxies
included in the composite range from 1.26$-$2.55, with a median
redshift of $z=2.1$.  Because the continuum emission varies
dramatically from source to source, we first remove continuum emission
before stacking.  For high signal-to-noise continuum ($>$10) we fit
the continuum using a third order polynomial, and for low
signal-to-noise continuum, we use a linear fit.  Continuum-subtracted
spectra are combined using an unweighted co-addition.  We test several
combination methods, including weighting individual spectra by the
strength of \ha\ emission, weighting by signal-to-noise of \ha,
weighting by the spectrum's inverse-variance (measured using each
spectrum's extracted 1D flux error array), and co-adding the spectra
without special weighting.  All combinations produce similar spectra,
but the unweighted spectrum, scaled by flux, produces the most
uniform, high signal-to-noise result.  We check to ensure the final
spectrum is not dominated by the brightest sources; the brightest
\ha\ source, 850.44, contributes 16\%\ to the final stack.  The median
contribution is 3.5\%, and the minimum contributor (faintest)
contributes 1\%.

\begin{figure}
\centering
\includegraphics[width=0.75\columnwidth]{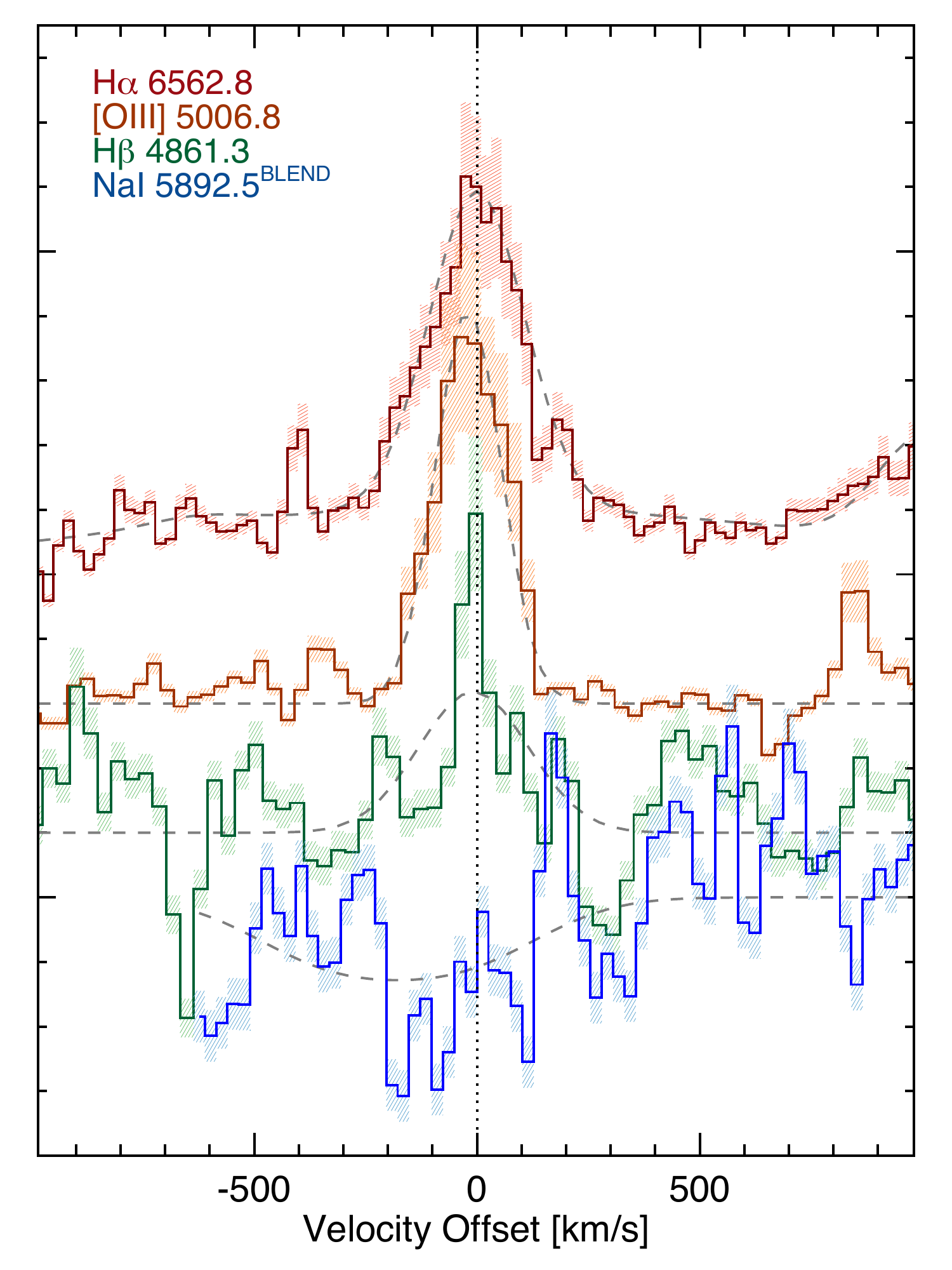}
\caption{ Velocity offset and width comparison across different line
  tracers in the stacked spectrum, including \ha\ (from which the
  systemic redshifts are inferred), \oiii, \hb, and finally Na\,B
  absorption.  The Na\,B absorption is slightly blueshifted,
  indicative of winds.}
\label{fig:stackzoom}
\end{figure}

The resulting rest-frame optical composite for DSFGs is shown in
Figure~\ref{fig:stackha}, extending from rest-frame wavelengths
$\sim$4300-7300\AA.  We compare this fit to a similar composite
derived in \citet{swinbank04a} for 23 850\um-selected DSFGs
(i.e. SMGs).  Given the spectral coverage of Keck NIRSPEC which was
used for the SMG sample the Swinbank \etal\ composite covered a much
narrower rest-frame wavelength range.  We simultaneously fit \ha\ and
\nii\ emission in the range 6550--6600\AA\ using four co-added Gaussian
fits, with the narrow band component of \ha\ and the two
\nii\ features at a fixed width. The \ha\ line is best fit with both a
broad and narrow component, rather than a single fit.  The narrow
component has a FWHM of $\Delta v=253\pm6$\,\kms\ (comparable to the
results of Swinbank \etal), while the broad component is $\Delta
v=2000\pm80$\,\kms.  The narrow-to-broad flux ratio is 0.61$\pm$0.03,
nearly identical to the ratio found in the Swinbank \etal\ composite.

The \nii/\ha\ ratio for the composite is 0.33$\pm$0.02, consistent
with a star-formation origin.  We detect both \sii\ features at
6716\AA\ and 6731\AA, and measure the \sii/\ha\ ratio at
0.15$\pm$0.01, consistent with an H\,{\sc ii} region origin
\citep{veilleux87a}.  We also use the ratio of the two sulfur lines to
infer the mean electron density in DSFGs.  A ratio of
\sii6716/\sii6731 = 0.86$\pm$0.12 implies an electron density of
$n_{e}\sim$900$\pm$300\,cm$^{-3}$. This density is well above the
$\simlt$10\,cm$^{-3}$ low density regime \citep*[see Figure 5.8
  of][]{osterbrock06a} of many lower luminosity galaxies, and even
$\sim$200\,cm$^{-3}$ densities seen in high-$z$ star-forming galaxies
\citep{strom16a}, implying a dense interstellar medium (ISM)
consistent with DSFGs' relatively compact, high SFR-density, and high
gas-density characteristics.
The measured ratio of \ha\ to \hb\ emission in the composite, i.e. the
Balmer decrement, is 21.0$\pm$2.4 which translates to an
  $A_{V}=6.8\pm0.5$.  This is, of course, substantially higher than
the theoretically expected \ha/\hb\,=2.86 from case B recombination,
and the dramatic difference is directly attributable to the extreme
obscuration present in the DSFG population.

The composite spectrum also shows notable Na absorption blueshifted by
$-$200$\pm$30\,\kms, indicative of gas outflows. This blueshift is
consistent with the measured $-$240$\pm$50\,\kms\ blueshift observed in
Mg\,II and Fe\,II features in slightly lower redshift DSFGs
\citep{banerji11a}, indicative of large-scale outflowing interstellar
gas and consistent with the momentum-driven wind model.  This is also
broadly consistent with the local relation of SFR to ISM outflow
velocity from \citet{martin05a}.

In Figure~\ref{fig:stackzoom} we compare the line profiles of \ha,
\oiii, \hb\ and Na\,I from the composite spectrum.  The measured
\oiii\ 5007\AA\ width is notably narrower than the width of \ha, at
74.0$\pm$3.7\,\kms, though the \hb\ line width is consistent with the
\ha\ width. The \oiii/\hb\ ratio is 5.17$\pm$0.78.  As we show in the
next sub-section, we use both the \nii/\ha\ and \oiii/\hb\ ratios to
place our composite spectrum, as well as individual DSFGs with bright
features, on a nebular emission line diagram, and compare it to
literature sources at low and high-redshift.

%

\section{AGN Content}\label{sec:agn}

AGN can significantly impact rest-frame optical line diagnostics.
Here we summarize the contribution of AGN to the spectroscopically
confirmed population of DSFGs.  Only four of the 31
spectroscopically-confirmed sources are X-ray detected: 450.58, 850.36,
850.89 and m450.133.  Their integrated rest-frame 0.5-8\,keV
luminosities are 8.0$\times$10$^{43}$\,erg\,s$^{-1}$,
3.1$\times$10$^{41}$\,erg\,s$^{-1}$,
2.0$\times$10$^{44}$\,erg\,s$^{-1}$ and
1.1$\times$10$^{44}$\,erg\,s$^{-1}$, respectively.  The low luminosity
for 850.36 (at $z=0.224$) is consistent with a star-formation origin,
while the other sources are consistent with luminous AGN.  Sources not
detected in X-ray data but showing signatures of bright AGN include
450.09, discussed in \citet{casey15a} as being radio-loud, just under
the classic FR\,II classification luminosity.  As shown in
Figure~\ref{fig:magphys}, 450.09 is one of two sources with a clear
powerlaw through mid-infrared wavelengths, consistent with AGN-heated
torus dust.  The other is 850.59.  Together, these rather clear AGN
indicators in the X-ray and radio account for $\sim$10\%\ of the
confirmed DSFG sample, a noticeably lower fraction than found at higher
luminosities.

While most of our DSFGs lack obvious AGN signatures, some of the
remaining DSFGs in our sample may contain AGN that contribute at a
lower level to a galaxy's bolometric luminosity.  While we do have 20
DSFGs for which an N$_2\equiv$\nii/\ha\ ratio is measurable, nearly
half the sample (9) have $\log($\nii/\ha$)>-0.5$, consistent with
solar or super-solar metallicity, shock-heating or AGN.  As revealed
from our composite spectrum in Figure~\ref{fig:stackha}, there is an
underlying broad \ha\ line profile with 2000\,\kms\ width confirming
the presence of AGN, albeit at a level which does not dominate the
bolometric luminosity of our sources, but further demonstrates the
link between luminous starbursts and their nuclear activity.

\section{Nebular Emission Diagnostics}\label{sec:bpt}

Five DSFGs have sufficiently high-S/N spectra with independent
detections of \ha, \nii, \oiii\ and \hb\ to place in the context of
the strong-line Baldwin-Phillips-Terlevich (BPT) classification
\citep*{baldwin81a}, used to infer the dominant mode of excitation for
strong nebular emission in both low and high redshift galaxies.  While
this sample is insufficient to independently fit as a population, it
does provide some context for a handful of typical high-$z$ DSFGs, and
provides some promise that future MOSFIRE surveys of larger DSFG
samples could become a useful technique for understanding the galaxies'
internal physical drivers.  This is particularly useful in
understanding the origins of H{\sc ii} region ionization, either from
AGN or UV radiation from young stars \citep{kewley01a}.

\begin{figure}
\includegraphics[width=0.99\columnwidth]{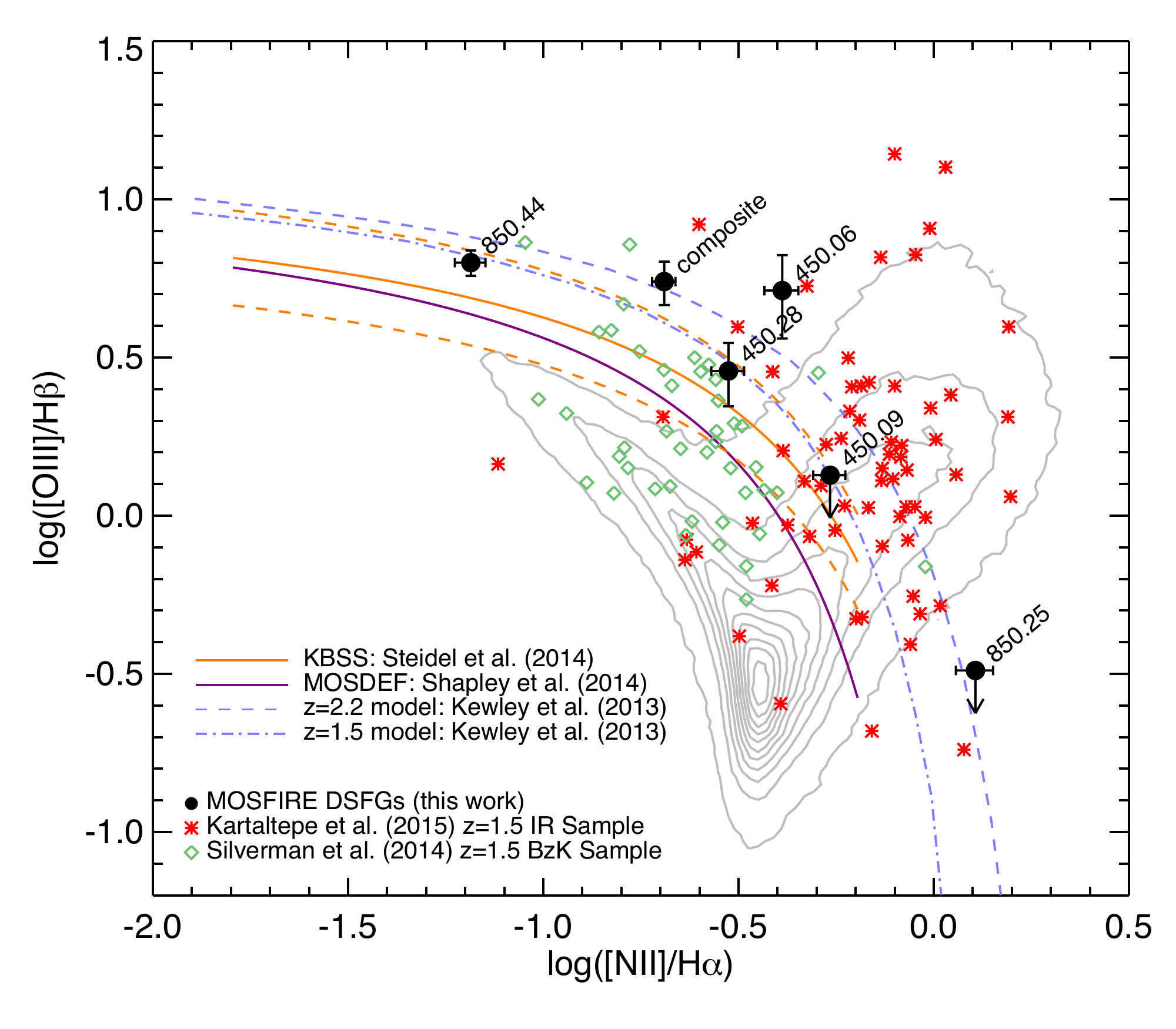}
\caption{The BPT diagram for the DSFGs of this sample and the
  composite spectrum (so labeled).  In comparison we show the $z=0$
  SDSS cloud \citep[gray;][]{kewley01a}, the $z=1.5$ (dot-dashed) and
  $z=2.2$ (dashed) model predictions of \citet{kewley13a} in light
  blue, the best-fit star-formation track from KBSS
  \citep[orange;][]{steidel14a} and MOSDEF
  \citep[purple;][]{shapley14a}, and $z\sim1.5$ BzK galaxies
  \citep[green diamonds;][]{silverman15a} and $z\sim1.5$ {\it
    Herschel}-PACS DSFGs \citep[red stars;][]{kartaltepe15a}.}
\label{fig:bpt}
\end{figure}

Figure~\ref{fig:bpt} shows our DSFGs in context with other high-$z$
samples.  The largest $z\sim2$ galaxy samples with these nebular line
diagnostics are drawn from the Keck Baryonic Sky Survey
\citep[KBSS;][]{steidel14a,strom16a} and from the MOSFIRE Deep
Evolution Field \citep[MOSDEF;][]{shapley14a}.  Both populations show
elevated \oiii/\hb\ ratios, with differences between the star-forming
tracks due to sample selections: in the rest-frame UV for KBSS vs. in
the observed near-IR for MOSDEF.  We also draw comparisons with
$z\sim1.5$ BzK-selected star-formers \citep{silverman15a} and {\it
  Herschel}-PACS selected DSFGs \citep{kartaltepe15a}.  Our composite
spectrum and 3/5 individual sources are on the upper envelope of
\citet{steidel14a} sample fits and the \citet{kewley13a} $z=2.2$
model.  While the \citeauthor{kartaltepe15a} work presents the largest
sample of DSFGs' with these optical line characteristics to-date, our
DSFGs here are more in-line with $z=1.5$ BzK galaxies and seem to
probe a different regime.  Indeed, AGN are a significant contributor
to the IR luminosity for {\it Herschel}-PACS samples, and as our
selection is at 450\um/850\um, this work is more sensitive to pure
starbursts.  The general observation of higher \oiii/\hb\ ratios, in
excess of those observed by \citet{steidel14a} and \citet{shapley14a}
is attributed to even harder ionizing radiation fields in these
starbursting systems with low metallicity.

As discussed in \citet{masters16a} and \citet{strom16a}, the N/O ratio
seems to be rather fundamental, holding a tight relation with
mass-metallicity.  At high stellar masses (which our DSFGs are) we
observe an even higher \nii/\ha\ ratio at a given \oiii/\hb.  Our
results are also consistent with those of \citet{stanway14a} who argue
that high \oiii/\hb\ ratios can be accounted for with binary stellar
evolution models in short-lived ($\sim$100\,Myr) starbursts.

%

\section{Extinction in DSFGs}\label{sec:ha}

\begin{figure}
\includegraphics[width=0.99\columnwidth]{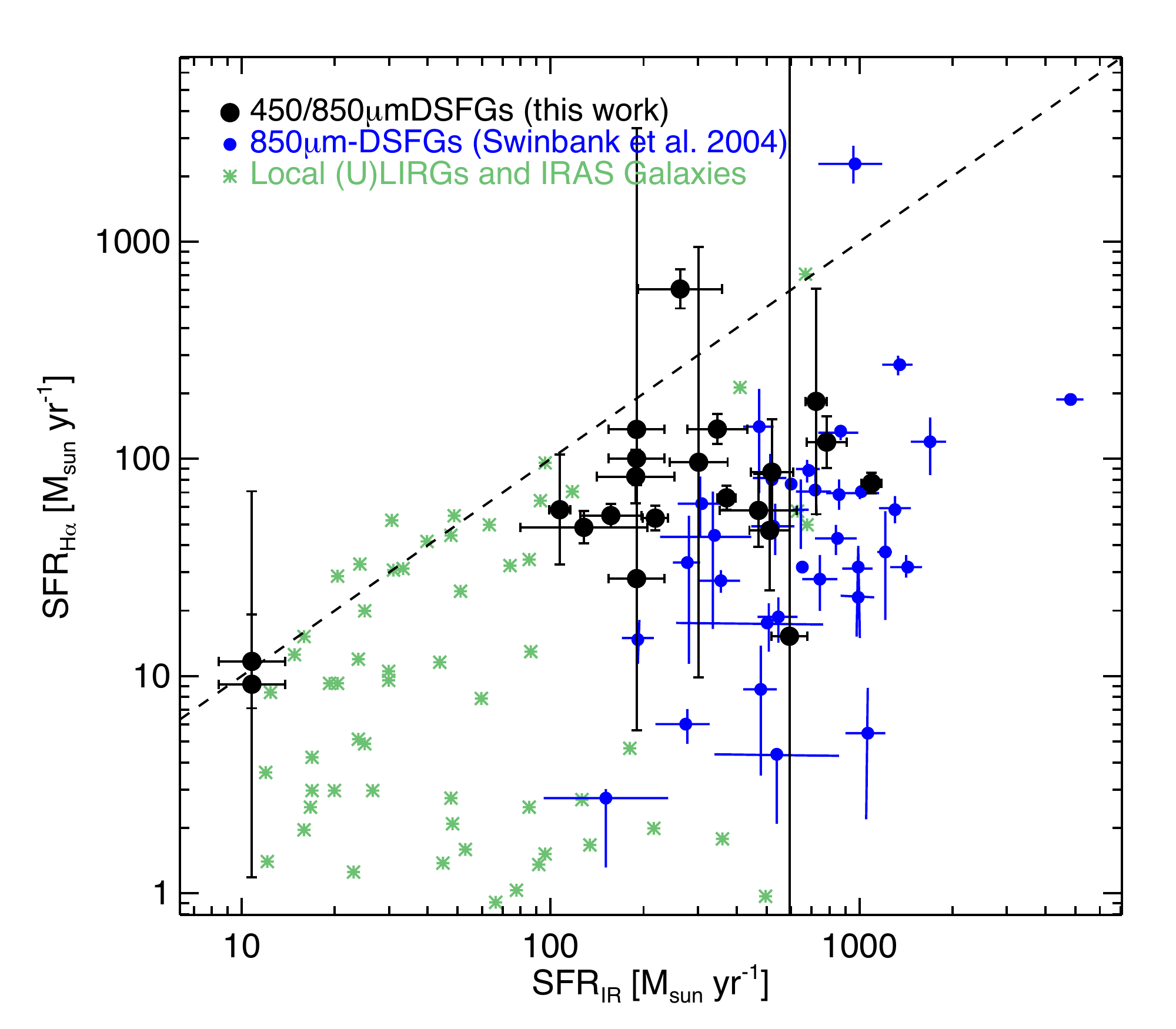}
\caption{\ha\ star-formation rates against IR star-formation rates for
  both local and high-redshift sources.  The \citet{swinbank04a} DSFGs
  (blue points) sit at slightly higher luminosities than the sample of
  this paper (black points).  Both high-$z$ DSFG samples exhibit
  substantial extinction, rendering the \ha\ luminosity-to-SFR scaling
  inaccurate.  This is reminiscent of more nearby galaxies
  \citep{franceschini03a,flores04a}.  While the \ha\ luminosities of
  the Swinbank \etal\ sample underestimate the total SFRs by factors
  of $\sim$14, our lower luminosity sample only under-predicts the
  total SFR by factors of $\sim$1.3.}
\label{fig:hasfr}
\end{figure}

\begin{figure}
\includegraphics[width=0.99\columnwidth]{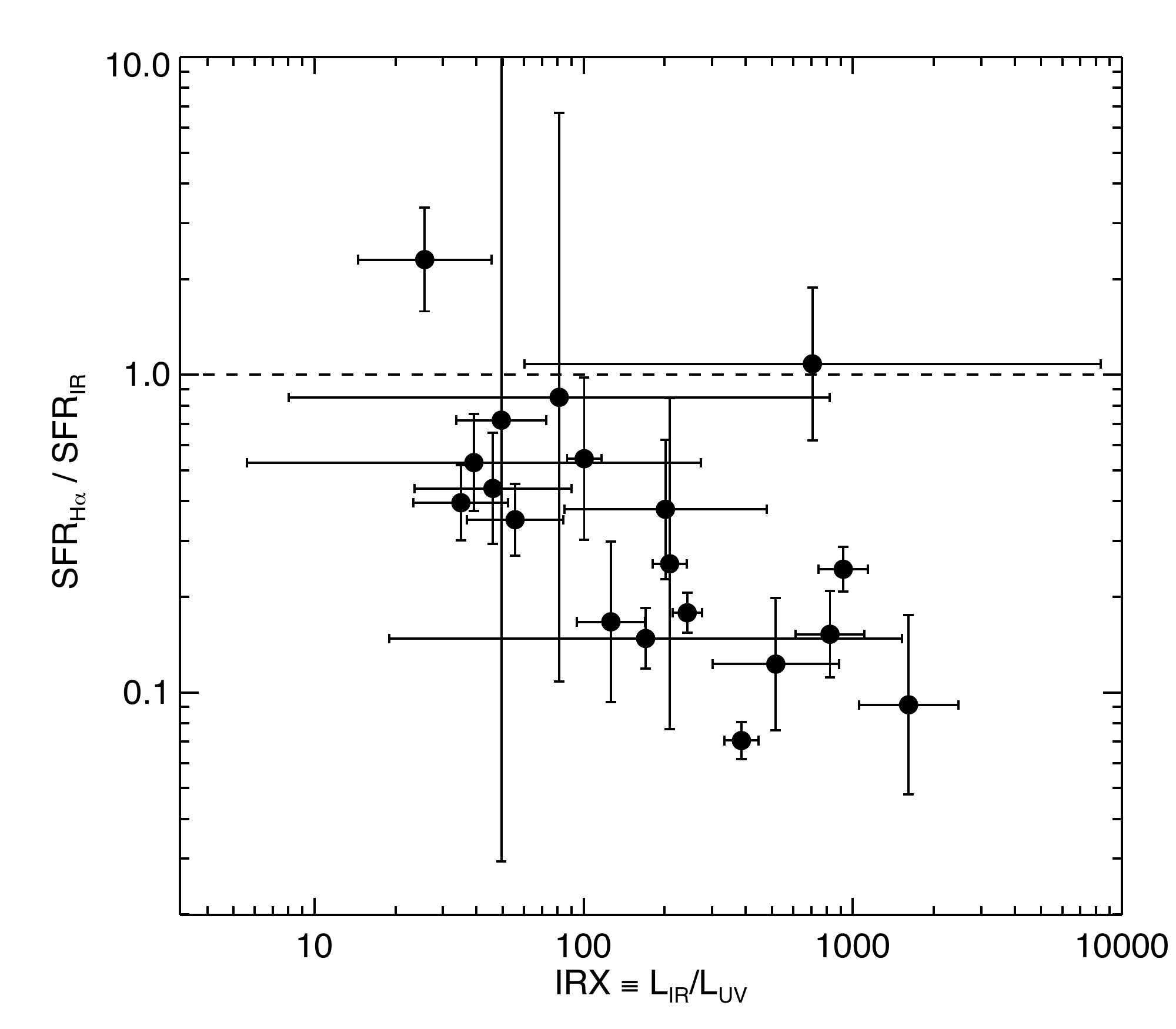}
\caption{The IRX ratio, or the ratio of IR to UV luminosity (a tracer
  of relative obscuration of a galaxy) against the ratio of measured
  star-formation rates from \ha\ and IR.  In these galaxies, IR can be
  viewed as the total SFR.  It is clear that at higher obscurations,
  the \ha\ SFR is less and less representative of the total SFR of the
  system, even though \ha\ emission itself is relatively unobscured
  compared to the rest-frame UV continuum emission, probed by the IRX
  measurement.}
\label{fig:irx_ha}
\end{figure}

DSFGs are known to be among the most heavily extinguished sources at
rest-frame ultraviolet and optical wavelengths, and here we explore
where this sample of DSFGs sit with respect to other DSFG samples in
terms of \ha\ extinction.  Figure~\ref{fig:hasfr} plots the integrated
infrared-based star-formation rate against the H$\alpha$ based
star-formation rate, as inferred directly from \ha\ line luminosity
\citep{kennicutt98a}.  Compared to the Swinbank \etal\ sample of
850\um-selected DSFGs, our sample is a bit less luminous at
far-infrared wavelengths, yet similarly bright (or brighter) in
\ha\ luminosity.  Swinbank \etal\ attribute the factor of 14$\pm$7
discrepancy between IR and \ha\ SFR indicators to obscuration of the
rest-frame optical, albeit not as significant as the much stronger
discrepancy between IR and rest-frame UV, which is typically a factor
of $\sim$120 \citep[IRX$\equiv L_{\rm IR}/L_{\rm
    UV}$;][]{chapman05a}. As our sample is, on average, less luminous
and less extreme than the Swinbank sample, the average \ha\ SFR
deficit we measure is only 1.3$\pm$0.1, in comparison.  In other
words, \ha\ in our sample is recovering about 77\%\ of the total SFR
of our DSFGs, while in the more luminous Swinbank \etal\ DSFG sample,
only $\sim$7\%\ is recovered.  This type of extinction is reminiscent
of similar effects seen at rest-frame UV wavelengths.  For example, in
\citet{casey14b} we showed that the  IRX ratio (or more specifically the
deviation from the IRX-$\beta$ relationship) is a strong function of
galaxies' total star-formation rates.  As SFR increases, attenuation
becomes more severe due to geometric effects decoupling UV and IR
emission.  In Figure~\ref{fig:irx_ha}, we show that the \ha\ SFR
deficit tracks the IRX ratio well, with the highest-obscuration UV
sources are those DSFGs whose \ha\ emission most dramatically
underestimates the total SFR.

\section{Conclusions}\label{sec:conclusions}

This paper has presented new spectroscopic observations of DSFGs in
the COSMOS field which were initially selected via their emission at
submillimeter wavelengths, at 450\um\ and 850\um.  Of 114 sources
initially targeted by both DEIMOS and MOSFIRE observations, we have
spectroscopically confirmed 31.  The vast majority of the sources
identified were through MOSFIRE near-infrared spectroscopy (where we
targeted 102 sources), with a few sources revealed by LRIS and DEIMOS
optical spectroscopy.

The vast majority (71/102) of our MOSFIRE spectroscopic targets did
not yield redshift identifications. 
From their photometric redshifts, we estimate about 60\%\ of these
failures are likely caused by the sources sitting outside the optimum
redshift range where bright emission lines are detectable in MOSFIRE
$H$- and $K$-bands, while the other 40\%\ are likely too obscured at
rest-frame optical wavelengths to be detected in, e.g., \ha\ emission.
Archival ALMA dust continuum data exist for seven sources from our
spectroscopic survey; five of the seven sources (71\%) were correctly
identified using multiwavelength counterparts.

Because the COSMOS field has some of the highest quality photometric
redshifts available thanks to 30$+$ bands of deep imaging, we assess
the quality of photometric redshifts in DSFGs, and conclude that half
are of high quality, $\Delta z/(1+z)<0.05$, the other half are poor
fits (no photometric redshift or $\Delta z/(1+z)\approx0.6$), which we
attribute to geometric decoupling of the galaxies' unobscured and
obscured emission.  Indeed, the sources whose photometric redshifts
fail catastrophically have the most significant offset between
rest-frame optical/near-infrared SED characteristics and
far-infrared/submm characteristics.

%

We have constructed a composite spectrum of 20 DSFGs in the rest-frame
optical spanning 4300--7300\AA, much wider wavelength coverage than
the previous DSFG composite spectra thanks to the sensitivity of the
MOSFIRE instrument and our deep spectroscopic observations in both $H$-
and $K$-bands.  We detect \ha, \nii, \sii, Na\,D, \oiii, and \hb\ in our
composite and conclude that the DSFGs, in aggregate, are
star-formation dominated as measured from their \nii/\ha\ ratio, have
relatively high electron densities $\sim$1000\,cm$^{-3}$, and
unsurprisingly are significantly obscured, with a Balmer decrement of
21.0$\pm$2.4.  A handful of individual sources are detected at
high-S/N across several nebular emission tracers to allow
characterization on the BPT diagram.  We find our sample of DSFGs more
skewed towards star-formation driven ionization rather than luminous
AGN, but with harder ionizing radiation fields than lower redshift
galaxies and lower-SFR galaxies at similarly high redshifts.

In line with the observation that our sample is, on average, less
extreme than previously studied samples of DSFGs with \ha\ observations, we find
that their \ha\ star-formation rates only underestimate the total SFR
of the system by a factor of 1.3$\pm$0.1, in contrast to much larger
factors $>$10.  The most extreme \ha\ SFR deficits align with the most
extreme IRX, or $L_{\rm IR}/L_{\rm UV}$, ratios.

Overall, this survey, like the ALESS survey \citep{danielson16a}, has
revealed that spectroscopic redshifts $-$ the classic ``bottleneck'' of
DSFG analysis $-$ are just as elusive today as they have been in the
past decade.  While newer, sensitive wide-bandwidth technology has
come online at long wavelengths in recent years (most notably the
Atacama Large Millimeter Array), there have not yet been large
spectroscopic programs pursued in the millimeter to ease pressure from
optical/near-infrared facilities in spectroscopically characterizing
DSFGs.  Until then, optical/near-infrared facilities will still be the
most efficient source of DSFG spectroscopy despite the population's
high obscuration.

\acknowledgements

This work was supported in part by a NASA Keck PI Data Award,
administered by the NASA Exoplanet Science Institute.  The data
presented herein were obtained at the W.M. Keck Observatory which is
operated as a scientific partnership among the California Institute of
Technology, the University of California and the National Aeronautics
and Space Administration.  The Observatory was made possible by the
generous financial support of the W.M. Keck Foundation.  The authors
wish to recognize and acknowledge the very significant cultural role
and reverence that the summit of Maunakea has always had within the
indigenous Hawaiian community.  We are most fortunate to have the
opportunity to conduct observations from this mountain.  

This work was performed in part at the Aspen Center for Physics, which
is supported by National Science Foundation grant PHY-1066293.  
This paper makes use of the following ALMA data:
ADS/JAO.ALMA\#2013.1.00118.S, 2013.1.00151.S, and 2011.1.00539.S. ALMA
is a partnership of ESO (representing its member states), NSF (USA)
and NINS (Japan), together with NRC (Canada), NSC and ASIAA (Taiwan),
and KASI (Republic of Korea), in cooperation with the Republic of
Chile. The Joint ALMA Observatory is operated by ESO, AUI/NRAO and
NAOJ. The National Radio Astronomy Observatory is a facility of the
National Science Foundation operated under cooperative agreement by
Associated Universities, Inc

COSMOS is based on observations with the NASA/ESA {\it Hubble Space
  Telescope}, obtained at the Space Telescope Science Institute, which
is operated by AURA Inc, under NASA contract NAS 5-26555; also based
on data collected at: the Subaru Telescope, which is operated by the
National Astronomical Observatory of Japan; the XMM-Newton, an ESA
science mission with instruments and contributions directly funded by
ESA Member States and NASA; the European Southern Observatory, Chile;
Kitt Peak National Observatory, Cerro Tololo Inter-American
Observatory, and the National Optical Astronomy Observatory, which are
operated by the Association of Universities for Research in Astronomy,
Inc. (AURA) under cooperative agreement with the National Science
Foundation; the National Radio Astronomy Observatory which is a
facility of the National Science Foundation operated under cooperative
agreement by Associated Universities, Inc; and the
Canada-France-Hawaii Telescope operated by the National Research
Council of Canada, the Centre National de la Recherche Scientifique de
France and the University of Hawaii.
CMC thanks the University of Texas at Austin, College of Natural
Science for support. AC is supported by NSF ST-1313319 and NASA
NNX16AF39G and NNX16AF38G.

\bibliography{caitlin-bibdesk}

\end{document}